\documentstyle[12pt,aasms4]{article}
\def\la{\mbox{\raisebox{-0.1ex}{$\scriptscriptstyle \stackrel{<}{\sim}$\,}}}
\def\ga{\mbox{\raisebox{-0.1ex}{$\scriptscriptstyle \stackrel{>}{\sim}$\,}}}

\newcommand{\Fig}{\mbox{\sc Fig. }}

\newcommand{\cone}{\mbox{$ { {\rm C_1 } } $\,} }
\newcommand{\ctwo}{\mbox{$ { {\rm C_2 } } $\,} }
\newcommand{\cthree}{\mbox{$ { {\rm C_3 } } $\,} }

\newcommand{\nd}{\mbox{$ \nu _d $ }}
\newcommand{\td}{\mbox{$ \tau _d $~}}
\newcommand{\ndi}{\mbox{ $ \nu _{d,i} $ }}
\newcommand{\vsmod}{\mbox{ $ \sigma ^2 _{mod} $}}
\newcommand{\vsest}{\mbox{ $ \sigma ^2 _{est} $}}
\newcommand{\vscal}{\mbox{ $ \sigma ^2 _{cal} $}}
\newcommand{\smod}{\mbox{ $ \sigma _{mod} $}}
\newcommand{\sest}{\mbox{ $ \sigma _{est} $}}
\newcommand{\scal}{\mbox{ $ \sigma _{cal} $}}
\newcommand{\smeas}{\mbox{$\sigma _{meas}$\,}}
\newcommand{\tref}{\mbox{ $ {\rm \tau _{ref} } $ }}
\newcommand{\nref}{\mbox{ $ {\rm N _{ref} } $ }}

\newcommand{\mb}{\mbox{${ m_b }$ }}
\newcommand{\avmb}{\mbox{${ \langle m_b \rangle }$ }}
\newcommand{\avmbc}{\mbox{${ \langle m_{b,c} \rangle }$ }}
\newcommand{\avmt}{\mbox{${ \langle m_t \rangle }$ }}
\newcommand{\avmr}{\mbox{${ \langle m_r \rangle }$ }}
\newcommand{\mbc}{\mbox{${ m_{b,c} }$ }}
\newcommand{\mbcriss}{\mbox{${ m_{b,c(riss)} }$ }}
\newcommand{\mt}{\mbox{${ m_t }$\,}}
\newcommand{\mr}{\mbox{${ m_r }$\,}}
\newcommand{\mlin}{\mbox{${ m_{lin} }$\,}}
\newcommand{\mrpol}{\mbox{${ m_{r,pol} }$\,}}
\newcommand{\mrriss}{\mbox{${ m_{r,riss} }$ }}
\newcommand{\mriss}{\mbox{${ m_{riss} }$ }}
\newcommand{\mnoise}{\mbox{${ m_{noise} }$\,}}
\newcommand{\mmeas}{\mbox{${ m_{meas} }$\,}}
\newcommand{\deltvobs}{\mbox{ $ {\rm { \delta t _{vobs} } } $\,}}
\newcommand{\deltvirr}{\mbox{ $ {\rm { \delta t _{virr} } } $\,}}

\newcommand{\dnt}{\mbox{ $ d \nu / d t $ }}

\newcommand{\dtn}{\mbox{ $ d t / d \nu $\,}}
\newcommand{\avdtn}{\mbox{ $ \langle d t / d \nu \rangle $\,}}
\newcommand{\rmsdtn}{\mbox{ $ \delta ( d t / d \nu ) $\,}}
\newcommand{\ndc}{\mbox{ $ \nu _{d_c} $\,}}
\newcommand{\ndci}{\mbox{$\nu _{d_c,i}$\,}}
\newcommand{\tdi}{\mbox{$\tau _{d,i}$\,}}
\newcommand{\ndg}{\mbox{ $ \nu _{d,g} $\,}}
\newcommand{\tdg}{\mbox{ $ \tau _{d,g} $\,}}

\newcommand{\avnd}{\mbox{$ \langle \nu _d \rangle $\,}}
\newcommand{\avndc}{\mbox{$ \langle \nu _{d_c} \rangle $\,}}
\newcommand{\avtd}{\mbox{ $ \langle \tau _d \rangle $\,}}

\newcommand{\avfd}{\mbox{ $  \langle  \rm F \rangle  $\,}}

\newcommand{\delDM}{\mbox{ $ {\rm \delta DM } $\,}}
\newcommand{\diff}{\mbox{$ {\rm \theta _{diff} } $ }}
\newcommand{\diffi}{\mbox{$ {\rm \theta _{diff,i} } $ }}
\newcommand{\avdiff}{\mbox{${\rm \langle \theta _{diff} \rangle}$~}}
\newcommand{\refr}{\mbox{${\rm \theta _{ref} }$\,}}
\newcommand{\Refr}{\mbox{ $ {\rm \Theta _{ref} } $ }}

\newcommand{\refri}{\mbox{ $ {\rm \theta _{ref,i} } $ }}
\newcommand{\refkol}{\mbox{$ {\rm \delta \theta _{r,kol} } $ }}
\newcommand{\refmean}{\mbox{$ {\rm \langle \theta _{ref} \rangle } $ }}
\newcommand{\avref}{\mbox{$ {\rm \langle \theta _{ref} \rangle } $ }}
\newcommand{\varref}{\mbox{$ {\rm \langle \theta _{ref} ^2 \rangle } $ }}
\newcommand{\varRef}{\mbox{$ {\rm \langle \Theta _{ref} ^2 \rangle } $ }}

\newcommand{\refrms}{\mbox{$ {\rm \delta \theta _{ref} } $ }}
\newcommand{\rmsref}{\mbox{$ {\rm \delta \theta _{ref} } $ }}
\newcommand{\Rmsref}{\mbox{$ {\rm \delta \Theta _{ref} } $ }}

\newcommand{\avbeta}{\mbox{$ {\rm \langle \beta \rangle } $ }}
\newcommand{\avalpha}{\mbox{$ {\rm \langle \alpha \rangle } $ }}
\newcommand{\betasteep}{\mbox{$ {\rm \langle \beta _{steep} \rangle } $ }}

\newcommand{\phiref}{\mbox{$ {\rm \phi _{ref} } $\,}}

\newcommand{\Dphi}{\mbox{$ {\rm D _{\phi} }$ }}
\newcommand{\kdiff}{\mbox{$ {\rm \kappa _{diff} }$\,}}
\newcommand{\kref}{\mbox{$ {\rm \kappa _{ref} }$\,}}
\newcommand{\krefalp}{\mbox{$ {\rm \kappa ^{-\alpha} _{ref} }$\,}}
\newcommand{\kdiffalp}{\mbox{$ {\rm \kappa ^{-\alpha} _{diff} }$\,}}
\newcommand{\kinn}{\mbox{$ {\rm \kappa _{inn} }$\,}}
\newcommand{\kinnsq}{\mbox{$ {\rm \kappa _{inn} ^2 }$\,}}
\newcommand{\koutsq}{\mbox{$ {\rm \kappa _{out} ^2 }$\,}}
\newcommand{\kout}{\mbox{$ {\rm \kappa _{out} }$\,}}
\newcommand{\pdiff}{\mbox{$ {\rm P _{diff} }$~}}
\newcommand{\pref}{\mbox{$ {\rm P _{ref} }$~}}
\newcommand{\sref}{\mbox{${\rm  s _{ref} }$ }}

\newcommand{\sdiff}{\mbox{${\rm s_{diff}}$ }}
\newcommand{\sinn}{\mbox{${\rm s_{inn}}$ }}
\newcommand{\sout}{\mbox{${\rm s_{out}}$ }}
\newcommand{\so}{\mbox{${\rm s_o}$ }}
\newcommand{\cng}{\mbox{$ C_n^2 $}}
\newcommand{\cn}{\mbox{${\rm C_n^2}$\,}}
\newcommand{\cN}{\mbox{${\rm C_N^2}$\,}}
\newcommand{\dmu}{$ {\rm pc ~ cm ^ {-3} } $ }
\newcommand{\rotu}{$ {\rm rad ~ m ^ {-2} } $ }
\newcommand{\cmc}{$ {\rm cm ^ {-3} } $ }
\newcommand{\dtnu}{$ {\rm sec ~ kHz ^ {-1} } $}
\newcommand{\velu}{\mbox{$ {\rm km ~ sec ^ {-1} } $}}
\newcommand{\nep}{\mbox{$ {\rm N_{ep} } $ }}
\newcommand{\Nele}{\mbox{$ {\rm N_e } $\,}}
\newcommand{\nele}{\mbox{$ n_e $\,}}
\newcommand{\avne}{\mbox{$ {\rm \langle n_e \rangle } $\,}}
\newcommand{\viss}{\mbox{$ V_{iss} $\,}}
\newcommand{\vissc}{\mbox{$ V_{iss,c} $ }}
\newcommand{\vobs}{\mbox{$ V_{obs} $ }}

\newcommand{\delvep}{\mbox{ $ \Delta V_{obs _{\bot }} $}}
\newcommand{\virr}{\mbox{$ V_{irr} $ }}
\newcommand{\vissi}{\mbox{$ V_{iss,i} $ }}
\newcommand{\avvissc}{\mbox{$ \langle V_{iss,c} \rangle $ }}
\newcommand{\tsp}{\mbox {$ {\rm T_{sp} } $ }}
\newcommand{\fobs}{\mbox{$ f_{obs} $\,}}
\newcommand{\fobsGHz}{\mbox{$ f_{obs[{\rm GHz}]} $\,}}
\newcommand{\Bobs}{\mbox{$ {\rm B_{obs} } $\,}}
\newcommand{\Tpulse}{\mbox{$ {\rm \tau _{pulse} } $\,}}
\newcommand{\wavobs}{\mbox{${\rm \lambda _{obs}}$ }}
\newcommand{\wavobstwo}{\mbox{$ {\rm \lambda ^2 _{obs} } $}}

\newcommand{\kolind}{\mbox{$ { 11 \over 3 } $ }}
\newcommand{\ald}{\mbox{$ \alpha _1 $}}

\newcommand{\normb}{\mbox{$ {\rm PSR ~ B0823+26(II) } $  }}
\newcommand{\egta}{\mbox{$ {\rm PSR ~ B0834+06(I) } $  }}
\newcommand{\egtb}{\mbox{$ {\rm PSR ~ B0834+06(II) } $  }}
\newcommand{\egtc}{\mbox{$ {\rm PSR ~ B0834+06(III) } $  }}
\newcommand{\egtd}{\mbox{$ {\rm PSR ~ B0834+06(IV) } $  }}
\newcommand{\eleva}{\mbox{$ {\rm PSR ~ B1133+16(I) } $  }}
\newcommand{\elevb}{\mbox{$ {\rm PSR ~ B1133+16(II) } $  }}

\newcommand{\ninea}{\mbox{$ {\rm PSR ~ B1919+21(I) } $  }}
\newcommand{\nineb}{\mbox{$ {\rm PSR ~ B1919+21(II) } $  }}
\begin{document}

%%%%%%%%%%%%%%%%%%%%%%%%%%%%%%%%%%%%%%%%%%%%%%%%%%%%%%%%%%%%%%%%%%%%%%%%%%%%%%%%%%%%%%%%%%%%%%%%%%%%%%%%%%%

\title{\LARGE\bf Long-Term Scintillation Studies of Pulsars: \\ 
II. Refractive Effects and the Spectrum of Plasma Density Fluctuations}

\vspace{5.0cm}
 
\author{\bf N. D. Ramesh Bhat\footnote{send preprint requests to $ bhatnd@ncra.tifr.res.in $}, 
Yashwant Gupta, and A. Pramesh Rao }

%\affil{National Centre for Radio Astrophysics, Tata Institute of Fundamental Research, \\
%Post Bag 3, Ganeshkhind, Pune - 411 007, India}

\begin{center}
{\normalsize National Centre for Radio Astrophysics, Tata Institute of Fundamental Research, \\
Post Bag 3, Ganeshkhind, Pune - 411 007, India}
\end{center}

\vspace{5.0cm}
 
\begin{center} {\normalsize\bf Accepted for publication in The Astrophysical Journal} \end{center}

%%%%%%%%%%%%%%%%%%%%%%%%%%%%%%%%%%%%%%%%%%%%%%%%%%%%%%%%%%%%%%%%%%%%%%%%%%%%%%%%%%%%%%%%%%%%%%%%%%%%%%%%%%%

\begin{abstract}

Refractive scintillation effects in pulsars are powerful techniques for discriminating 
between different models proposed for the electron density fluctuation spectrum in the interstellar medium.
Data from our long-term scintillation study of eighteen pulsars in the dispersion measure range 
3$-$35 \dmu (Paper I) are used to investigate two important observable effects of
refractive scintillation, {\it viz.} (i) modulations of diffractive scintillation observables and flux 
density, and (ii) drifting bands in dynamic scintillation spectra. 
Our data provide simultaneous measurements of decorrelation bandwidth, scintillation time scale, flux
density and drift rate of patterns.
The observed modulations of the first three are compared with the available theoretical predictions, 
and constraints are placed on the power spectrum of plasma density fluctuations.
The measured modulation indices are found to be larger than predicted by a Kolmogorov 
form of density spectrum. 
The properties of the drift rate of patterns along with the diffractive scintillation parameters 
have been used to independently estimate the slope of the density power spectrum,
which is found to be consistent with a Kolmogorov form for several pulsars.
The contradictory results from these two independent methods of constraining the electron density spectrum 
are not reconcilable with the simple theoretical models based on power-law forms of density spectrum.
Our observations show anomalous scintillation behaviour like persistent drifting bands for some pulsars.
This can be interpreted as an excess power in the low wavenumber range 
($ \sim $ $ 10^{-12} - 10^{-13} $ $ {\rm m ^{-1} } $) compared to the Kolmogorov
expectations, or the existence of localized density structures.
The results from our observations are discussed in combination with those from earlier studies in an
attempt to understand the overall nature of the density spectrum.
The emerging picture is a Kolmogorov-like spectrum ($ \alpha \ \approx \ \kolind $) in the wavenumber range
$ \sim 10^{-6} \ {\rm m^{-1} } $ to $ \sim 10^{-11} \ {\rm m^{-1} } $,
which either steepens or has a bump near $ \sim 10^{-12} - 10^{-13} \ {\rm m^{-1} }$.
The accumulated data also suggest the existence of discrete density structures along some lines of sight.
We also discuss the possible implications of our results for the theoretical models.

\end{abstract}

{\it Subject headings: }
{ISM:Structure -- Pulsars:General -- Radio continuum:ISM -- Turbulence -- Plasmas -- Interstellar:Matter}

%%%%%%%%%%%%%%%%%%%%%%%%%%%%%%%%%%%%%%%%%%%%%%%%%%%%%%%%%%%%%%%%%%%%%%%%%%%%%%%%%%%%%%%%%%%%%%%%%%%%%%%%%%%
%%%%%%%%%%%%%%%%%%%%%%	intro as revised on 26-08-98

\section{Introduction }

The recognition of interstellar propagation effects in the long-term flux variations of pulsars (Sieber 1982)
led to the discovery of a new class of propagation effects in radio astronomy, known as Refractive 
Interstellar Scintillation (RISS) (Rickett, Coles \& Bourgois 1984). 
RISS is thought to arise from propagation through electron density inhomogeneities of spatial scales 
much larger than the Fresnel scale.
In contrast to Diffractive Interstellar Scintillation (DISS), 
applications of RISS extend beyond the field of pulsars, 
and it is thought to be the cause of low frequency variability (LFV) of compact extragalactic radio sources.
Theoretical treatments of the observable consequences of refractive scintillation can be found in
Cordes, Pidwerbetsky \& Lovelace (1986) and Romani, Narayan \& Blandford (1986)
(also see Rickett 1990; Narayan 1992; Hewish 1992 for recent reviews).
Pulsars are excellent objects for studying both DISS and RISS.
In addition to the familiar long-term (days to weeks at metre wavelengths) flux variations, refractive 
scintillation effects are expected to give rise to slow modulations (time scales $ \sim $ days to weeks)
of DISS observables such as decorrelation bandwidth (\nd) and scintillation time scale (\td).
Refraction due to large-scale density irregularities also gives rise to organized features such as sloping 
patterns in pulsar dynamic spectra (e.g. Smith \& Wright 1985; Roberts \& Ables 1982). 
RISS can also give rise to not so easily observable effects such as 
variations in the pulse arrival times and angular wandering of the apparent source positions. 
While DISS effects probe the density irregularities of spatial scales $ \sim $ $ 10^6 $ m to $ 10^8 $ m, 
RISS effects enable us to probe irregularities of much larger spatial scales ($ \sim $ $ 10^{10} $ m to 
$ 10^{12} $ m).
Therefore, simultaneous measurements of DISS and RISS properties are a powerful method of determining the 
nature of the electron density spectrum over a much wider range than that has been possible by DISS alone.

The density fluctuations can be characterized by their spatial wavenumber spectrum, for which there are 
various potential forms.
The spectrum can be taken as an `extended power-law' with the following general form (e.g. Rickett 1990).

\begin{equation}
{\rm P _{\delta n_e} } (\kappa) ~ = ~ \cn ~ 
\left( \kappa ^2 + \koutsq \right) ^{-(\alpha/2)} ~ 
{\rm exp} \left( - { \kappa ^2 \over \kinnsq } \right)
\end{equation}

\noindent
where $ \kappa $ is the spatial wavenumber, inversely related to the length scale $s$ 
(we use the relation $ \kappa = 1 / s $, in accordance with the convention of Armstrong, Rickett \& Spangler 1995), 
\kinn and \kout correspond to the inner and the outer cut-offs 
in scale sizes respectively.
The amplitude of the spectrum, \cn, is also known as the strength of turbulence.
The line of sight integral of equation (1) is a measure of the rms of electron density 
fluctuations, $ \delta n_e $.
The spectrum can be represented by a simpler form $ {\rm P _{\delta n_e} } (\kappa) ~ = ~ \cn ~ \kappa
^{-\alpha} $ in the range $ \kout \ll \kappa \ll \kinn $. 
The possibility that is most commonly discussed in the literature is the density fluctuations 
describable by a Kolmogorov spectrum, in which case $ \alpha = \kolind $.
We refer to this as hypothesis (I).
Two possible subsets of this hypothesis that may be relevant are: (IA) the cutoffs are not important
($ie,$ an inner scale much smaller than the smallest scale that influences the scintillation and 
an outer scale much larger than the largest scale sampled by the observations), and 
(IB) the spectrum is truncated at a large inner scale (say, intermediate between diffractive and 
refractive scales).
The second possibility is that the spectrum is a simple power-law (similar to IA), 
but with $ \alpha > \kolind $.
We will call this hypothesis (II).
The third possibility is that the density fluctuations are $ not $ describable by a simple power-law form 
over the entire range of spatial scales of interest: this can lead to a multi-component, ``piece-wise
power-law'' form (hypothesis IIIA) or a single power-law with an additional ``bump'' (hypothesis IIIB) 
at wavenumbers corresponding to density structures that are not part of the power-law process.
The fourth possibility is that the random medium (say, described by one of the abovesaid hypotheses) 
has deterministic structures superposed, in which case a power spectral description may not be adequate.
We will call this hypothesis (IV).
The exact form of the spectrum, especially the validity of a simple power-law description and 
the existence of inner and/or outer cutoff(s), is still a matter of research. 
In this paper, our main goal is to discriminate between the above hypotheses using the data from 
our long-term pulsar observations.

Observable effects of RISS are considered to be powerful techniques for discriminating between
different kinds of density spectra proposed for the ISM (e.g. Hewish, Wolszczan \& Graham 1985; Rickett 1990).
Refractive modulation characteristics such as depths of modulations of DISS observables
(\nd and \td), flux and drift slope are all thought to be highly sensitive to the spectral form 
(e.g. Romani et al. 1986).
But not all possible models indicated by hypotheses (I)$-$(IV) have been analyzed in detail. 
The $pure$ Kolmogorov form (type IA) is the best analyzed of all, and there are well defined
predictions for the magnitudes of observable effects.
The type II spectra have also been analyzed to some extent
(e.g. Blandford \& Narayan 1985; Romani et al. 1986). 
Three specific cases for which detailed analysis can be found in the literature are 
(i) $ \alpha = \kolind $, (ii) $ \alpha = 4 $ (`critical' spectrum), and (iii) $ \alpha = 4.3 $. 
The effects analyzed include depths of modulations and time scales of fluctuations for different
observables, refractive angle perturbations, scaling laws and cross-correlation properties between 
the fluctuations. 
While models based on hypothesis (IA) predict small-amplitude fluctuations of DISS observables and flux, 
and small refractive perturbations ($ \refr < \diff $), those based on hypothesis (II) 
can allow much larger fluctuations and large refractive angles ($ \refr \ \ga \ \diff $) 
(Romani et al. 1986; Hewish et al. 1985).
It has also been recognized that type IB spectrum, which has been suggested as an alternative to type 
II spectrum, can cause large-amplitude flux modulations (Coles et al. 1987; Goodman et al. 1987).
Effects (particularly perturbations on DISS observables) due to spectra grouped under hypothesis (III) 
have not been formally analyzed.
Observations such as periodicities seen in pulsar dynamic spectra and ``extreme scattering events'' (ESE)
seen towards some quasars 
and the millisecond pulsar PSR B1937+21 suggest that models based on hypothesis (IV) are also relevant
(e.g. Cordes \& Wolszczan 1986; Fiedler et al. 1987; Cognard et al. 1993).

Observationally, there are a number of ways of investigating the nature of the density fluctuation spectrum.
But there have been conflicting interpretations from different measurements.
The results from methods such as frequency scaling of decorrelation bandwidth (Cordes, Weisberg \& Boriakoff 1985), 
frequency drifts in dynamic spectra (e.g. Smith \& Wright 1985) and 
VLBI observations (Gwinn et al. 1988a, 1988b), are consistent with the Kolmogorov spectrum.
On the other hand, based on a study of dispersion measure (DM) variability, 
Phillips \& Wolszczan (1991) find the spectrum to be much steeper 
($\avalpha \approx 3.84 \pm 0.02$).
Backer et al. (1993) studied pulsars with comparatively larger DMs, and argue that DM variations are caused
by density fluctuations unrelated to those responsible for DISS. 
Contradictory results have come from flux monitoring observations of pulsars
(Kaspi \& Stinebring 1992; Gupta, Rickett \& Coles 1993; LaBrecque, Rankin \& Cordes 1994).
Observations of dynamic spectra at 408 MHz (for six pulsars) by Gupta, Rickett \& Lyne (1994) show 
the modulations of decorrelation bandwidth to be larger than the Kolmogorov expectations, 
but Cordes et al. (1990) find the modulations of PSR B1937+21 to be consistent with the 
Kolmogorov predictions.
Further, observations of quasi-periodic patterns in dynamic spectra 
(e.g. Cordes \& Wolszczan 1986) 
and detections of ESEs towards some quasars (Fiedler et al. 1987, 1994) go against the Kolmogorov 
expectations. 

Attempts have also been made to construct a {\it composite} spectrum by combining a variety of
scintillation measurements from different observations and for various pulsars and radio sources. 
The most recent study by Armstrong et al. (1995) finds the {\it average} spectral index to be 
$\approx$ 3.7 over the wavenumber range $ 10 ^{-13} - 10 ^{-8} \ {\rm m ^{-1}} $ for pulsars within 1 kpc.
They also find that, when combined with non-ISS measurements, the spectrum has an {\it approximately}
power-law form between $ 10 ^{-18} $ $ {\rm m ^{-1}} $ and $ 10 ^{-6} $ $ {\rm m ^{-1}} $.
Though this result is interesting, there is enough evidence in the literature suggesting that the
distribution of scattering material within a region of 1 kpc around the Sun is not homogeneous 
(Bhat, Gupta \& Rao 1997, 1998).
Furthermore, evidence for inadequacies of such a model can be seen in the composite spectrum of 
Armstrong et al.  (1995), in which the estimated power levels are discrepant from the power-law 
expectations at several places.
It is also possible that the nature of the density spectrum varies with the direction ($l,b$) and the
location in the Galaxy, but this aspect has not been systematically investigated so far.

Observational data from a systematic long-term study of diffractive and refractive scintillations of 
18 pulsars have been presented in our previous paper 
(Bhat, Rao \& Gupta 1998a, hereinafter referred to as Paper I).
These observations have yielded fairly accurate estimates of properties of observables such as 
decorrelation bandwidth (\nd), scintillation time scale (\td), flux density (F), 
and drift rate of patterns (\dtn).
Though our sample consists of mostly nearby pulsars (distance \la 1 kpc), there is a reasonably uniform 
coverage in $(l,b)$, DM and distance, and forms a more or less unbiased sample.
Furthermore, we have made simultaneous measurements of DISS and RISS properties from our data,
thereby reducing the possibility of observational bias.
In this paper, we study two important and easily observable effects that can be studied using our data, 
$viz,$ (i) modulations of DISS observables and flux density, and (ii) drifting of intensity patterns.
We examine the conformity of our data with the available quantitative predictions 
(for type IA and II spectra), and discuss the possible implications of our results 
for the form of the density spectrum.

The remainder of this paper is organized as follows.
In \S 2.1, we present our measurements of modulation indices (of \nd, \td and F) and compare them with the
theoretical predictions based on power-law models.
Then we describe the estimation of statistical properties of diffractive and refractive angles 
with which we estimate a slope parameter, indicative of the relative power level enhancement at
large scales (\S 2.2).
This is followed by a discussion on persistent drifting features seen for some pulsars (\S 2.3),
which suggest the presence of discrete structures, at least along some lines of sight.
In \S 3, we discuss the significance of various results from our observations and several others 
from the published literature, and study the implications for the nature of the density spectrum.
Our conclusions are presented in \S 4.

%%%%%%%%%%%%%%%%%%%%%%%%%%%%%%%%%%%%%%%%%%%%%%%%%%%%%%%%%%%%%%%%%%%%%%%%%%%%%%%%%%%%%%%%%%%%%%%%%%%%%%%%%%%%%
%\end{document}
%%%%%%%%%%%%%%%%%%%%%%%%%%%%%%%%%%%%%%%%%%%%%%%%%%%%%%%%%%%%%%%%%%%%%%%%%%%%%%%%%%%%%%%%%%%%%%%%%%%%%%%%%%%%%

\section{The Observational Data and Results}

The observational data used here come from an extensive series of scintillation measurements of 
18 pulsars in the DM range $3-35$ \dmu made using the Ooty Radio Telescope (ORT) 
at 327 MHz over a three-year period from 1993 to 1995.
Pulsars and the periods of observation are tabulated in columns (2) and (3) of Table 3.
The dynamic scintillation spectra of these pulsars were obtained at $ \sim $ 10$-$90 epochs spanning 
periods ranging from $ \sim $ 100 days to $ \sim $ 1000 days. 
Columns (4) and (5) in Table 3 give the number of epochs of observation (\nep) and the time span of
observation (\tsp) respectively.
The observations and the analysis methods used have been described in Paper I.

The observations were carried out in four well-separated sessions, each extending over a period of 
$ \sim $ 100 days, in which 6 to 8 pulsars were regularly monitored for their dynamic spectra at 
intervals of 1$-$2 days typically.
Four pulsars $-$ PSRs B0823+26, B0834+06, B1133+16 and B1919+21 $-$ were followed up for multiple
observing sessions: PSRs B0823+06 and B1919+21 for 2 sessions, PSR B1133+16 for 3 sessions, 
and PSR B0834+06 for all the 4 sessions.
The symbols I$-$IV, when attached along with these pulsar names, indicate the data from a particular session
(see Tables 1 and 2 of Paper I for more details) .
Most of the basic diffractive scintillation results have been presented in Paper I, including the time series 
of decorrelation bandwidth (\nd), scintillation time scale (\td), drift slope of intensity patterns
(\dtn) and pulsar flux density (F) (Figs. 4(a)$-$(x) of Paper I).
In this paper, we start with these time series and study their implications for the spectrum of the
electron density fluctuations in the ISM.

%%%%%%%%%%%%%%%%%%%%%%%%%%%%%%%%%%%%%%%%%%%%%%%%%%%%%%%%%%%%%%%%%%%%%%%%%%%%%%%%%%%%%%%%%%%%%%%%%%%5

\subsection{Refractive Modulations of Diffractive Scintillation Observables and Pulsar Flux Density}

\subsubsection{Predictions from Theoretical Models}

Due to refractive scintillation effects, measurable quantities such as \nd, \td, \dtn and F 
are expected to fluctuate with time.
Several authors have addressed the theory of refractive effects in pulsar scintillation 
(e.g. Cordes et al. 1986; Romani et al. 1986),
but only a few attempts have been made so far in measuring and verifying them.
The observable effects of RISS are thought to be highly sensitive to the form of the density spectrum, 
more specifically the relative power level enhancement at large scales compared to that at small scales. 
Romani et al. (1986) have worked out the theory for refractive effects due to simple power-law forms of 
density spectra with different spectral indices (covered under hypotheses IA and II of \S 1).
No explicit predictions are available at present for other kinds of spectra, such as those
covered under hypotheses (III) and (IV) of \S 1.

Considering  specific cases of $ \alpha $ = \kolind (Kolmogorov spectrum), $\alpha$ = 4 (`critical' 
spectrum), and $ \alpha $ = 4.3 (`steep' spectrum), Romani et al. (1986) find that the depth of 
modulation is the lowest for a Kolmogorov density spectrum, and increases for larger values of $ \alpha $.
In the simplest scattering geometry of a thin screen located midway between pulsar and observer,
the magnitude of fluctuations of the quantities \nd, \td and F will depend on
(i) the strength of scattering (\cn), (ii) observing wavelength (\wavobs), and (iii) 
distance to the pulsar (D), when the distribution of density irregularities follows a Kolmogorov form of 
power spectrum (hypothesis IA).
In contrast, the fluctuations are expected to be insensitive to these parameters for $ \alpha $ =4 and
$ \alpha $ = 4.3 spectra (hypothesis II).
Using the expressions given by Romani et al. (1986), for $ \alpha = \kolind $ spectrum,
the modulation indices of decorrelation bandwidth (\mb), scintillation time scale (\mt) 
and flux density (\mr) are given by

\begin{equation}
\mb ~ \approx ~ {\rm 9.8 \times 10^{-2} \times
\left( C_n^2 \right) ^{-0.2} ~ (\lambda _{obs,m}) ^{-0.6} ~ (D_{kpc}) ^{-0.4} }
\end{equation}

\begin{equation}
\mt ~ \approx ~ {\rm 4.8 \times 10^{-2} \times
\left( C_n^2 \right) ^{-0.2} ~ (\lambda _{obs,m}) ^{-0.6} ~ (D_{kpc}) ^{-0.37} }
\end{equation}

\begin{equation}
\mr ~ \approx ~ {\rm 1.2 \times 10^{-1} \times
( C_n^2 ) ^{-0.12} ~ (\lambda _{obs,m}) ^{-0.57} ~ (D_{kpc}) ^{-0.37} }
\end{equation}

\noindent
where $ C_n^2 $ is expressed in $ {\rm 10^{-4} ~ m ^{-20/3} } $.
Using these expressions, we obtain the predicted estimates for \mb, \mt and \mr for the pulsars
in our data set.  These are given in columns (4), (5) and (6) of Table 1.  For \cn values, we
use our results given in Paper I.
For pulsars with multiple observing sessions, the average of \cn estimates from different sessions is used.
Distance estimates used (given in column 3 of Table 1) are based on the the model for electron 
density distribution given by Taylor \& Cordes (1993), except for PSR B0823+26, for which we 
use the independent distance estimate from parallax measurements (Gwinn et al. 1986).
The approximate predictions for `steeper' spectra $-$ $ \alpha = 4 $ and $ \alpha = 4.3 $ $-$
are listed in Table 2 (see Romani et al. (1986) for details).

%%%%%%%%%%%%%%%%%%%%%%%%%%%%%%%%%%%%%%%%%%%%%%%%%%%%%%%%%%%%%%%%%%%%%%%%%%%%%%%%%%%%%%%%%%%%%%%%%%%%

\subsubsection{Results $-$ Modulation indices of \nd, \td and F}

From the time series presented in Figs. 4(a)$-$4(x) in Paper I, we estimate the modulation indices of
\nd, \td and F.
The rms fluctuations of any of these quantities (say \nd) is  given by

\begin{equation}
\Delta \nd \approx 
\left[ { 1 \over \nep } \sum _{i=1} ^{i=\nep} \left( \ndi - \avnd \right) ^2 \right] ^{0.5}
\end{equation}

\noindent
where \nep is the number of epochs of observations and \ndi denotes the measurement at $i^{th}$ epoch.
The modulation indices \mb, \mt and \mr, which are the fractional rms fluctuations of \nd, \td and F
respectively, are given by

\begin{equation}
\mb = { \Delta \nd \over \avnd }
\hspace{1.0cm}
\mt = { \Delta \td \over \avtd }
\hspace{1.0cm}
\mr = {\rm { \Delta F \over \avfd } }
\end{equation}

\noindent
where \avnd, \avtd and \avfd represent the average estimates.
We use the values \ndg and \tdg obtained from the
Global Auto-covariance Function (GACF) method described in Paper I
as estimators for \avnd and \avtd .
The mean flux density, \avfd, is computed from the time series directly.

Before seriously interpreting our results in terms of refractive scintillation, one needs to examine 
(i) the statistical reliability of the data, and
(ii) various possible reasons (other than RISS) for the fluctuations of the quantities.
In addition to the number of measurements (\nep), the number of refractive cycles spanned (\nref) also 
determines the statistical quality of our data. 
On the basis of \nep and \nref, we have divided our data into 3 broad categories, the details of which are
given in Appendix A and the results are summarized in columns (8) and (9) of Table 3.
The statistical reliability is best for data belonging to category ``A'' and reasonably good for those in
category ``B''.
The data which fall in either of the ``C'' categories are considered to be of poor statistical quality 
and are not taken seriously in our comparison with the predictions. 

The measured modulation indices range from 0.17 to 0.50 for \nd, 0.13 to 0.49 for \td, and 0.21 to 0.69 for F.
In general, values of \mt are comparatively smaller than those of \mb and \mr, and this is in qualitative 
agreement with the predictions given in Tables 1 and 2.
A visual examination of the time series $-$ Figs. 4(a)$-$(x) of Paper I $-$ shows that the observed 
fluctuations are generally random, but there are a few exceptions where some systematic trends can 
be seen over the time span of observation.
A closer inspection of the time series of \td measurements of PSR B2327$-$20 (Fig. 4(x) of Paper I) reveals 
a systematic downward trend where \td changes from $ \sim $ 1000 sec to $ \sim $ 200 sec over a 
span of 65 days. 
This is responsible for a substantially large value of \mt ($ 0.49 \pm 0.03 $) for this pulsar compared to 
the rest.
Excluding this outlier case, and also the data with poor statistical reliability, we find that \mt values 
range from 0.13 to 0.31.
For PSRs B1604$-$00 and B2016+28, some systematic trends are evident in their flux density time series.
The modulation indices are, however, not significantly higher than those of the rest.
Excluding the data with poor statistical quality, we find \mr values ranging from 0.23 to 0.57.
The global average modulation indices are 0.36 for \nd, 0.19 for \td and 0.45 for F.

There are various sources of errors and non-ISS effects that contribute to the observed modulation index.
These include 
(a) measurement noise, \smeas (applicable for all the 3 quantities),
(b) effect of variable Faraday rotation on flux density modulations, and
(c) effect of Earth's orbital motion on modulations of scintillation time scale.
A detailed treatment of these noise sources is presented in Appendix B and the estimates of their
contributions are summarized in Tables 4, 5 and 6.
The modulation indices due to the measurement noise are typically 0.1 and hence their contribution to the
measured modulation indices are only marginal for most of the data sets (see Table 4).
The noise-corrected modulation indices of \nd, \td and F are given in columns (6), (7) and (8) 
respectively of Table 4. 
Further, as can be seen from Tables 5 and 6, the effects of variable Faraday rotation and the Earth's 
orbital motion are significant only for a few pulsars.

%%%%%%%%%%%%%%%%%%%%%%%%%%%%%%%%%%%%%%%%%%%%%%%%%%%%%%%%%%%%%%%%%%%%%%%%%%%%%%%%%%%%%%%

\subsubsection{Comparison with the Predictions}

Taking into consideration various sources of errors and non-ISS effects discussed in Appendix B, 
and eliminating the data where such effects are found to be significant, 
we do a comparison study between the measured and predicted values of modulation indices.
Here we confine ourselves to the time series of 18 data sets where the statistical reliability is reasonable.
We also exclude the \td modulation indices of \normb (due to $ \mnoise \approx \mt $), PSR B$1604-00$
(due to $ \deltvobs \approx \mt $ and $ \deltvirr \approx \mt $) and PSR B$2327-20$ 
(due to the presence of a systematic trend in the time series) from the present comparison.
In general, we find most of the modulation indices (of \nd, \td and F) to be considerably larger than the 
Kolmogorov predictions given in Table 1, but there are a few exceptions.
The details are as follows:

(i) Flux density: 
There is no pulsar for which the measured value of \mr is in agreement with the predicted value given 
in Table 1.
While 12 of the measurements lie within the range between the predictions for $ \alpha = 4 $ and 
$ \alpha = 4.3 $ spectra (Table 2), the remaining 6 measurements are consistent with 
$ \kolind < \alpha < 4 $.

(ii) Decorrelation bandwidth:
There is only one measurement of \mb (\ninea) which agrees with the prediction given in Table 1.
Among the rest, 14 range between the predictions for $ \alpha $ = 4 and $ \alpha $ = 4.3
(Table 2), whilst 3 are between the predictions for $ \alpha = \kolind $ and $\alpha = 4$.

(iii) Scintillation time scale:
Only two measurements of \mt (\egtb and \egtc) agree with the predictions in Table 1.
Rest of the values range between the predictions for $\alpha = \kolind$ and $ \alpha $ = 4.3.

Though scattering from a thin screen is taken to be a good approximation in explaining diffractive 
scintillation phenomena, refractive effects may differ significantly depending on the scattering geometry 
considered.
There exist only a few theoretical treatments investigating refractive effects under more complex
scattering geometries such as one or more thick screens or an extended medium.
Romani et al. (1986) find that such scenarios will give rise to larger flux modulations
than those caused by a thin screen model.
According to the authors, if the scattering is uniformly distributed along the line-of-sight, 
the flux fluctuations will be larger by a factor 2.3 compared to the thin screen, 
in the case of a Kolmogorov form of spectrum ($ie,$ hypothesis IA of \S 1).
Coles et al. (1987) have also investigated the flux modulations for an extended medium, and their 
predicted modulation indices are comparable to those of Romani et al. (1986).
Similar estimates, however, do not exist for the rest of the observables of interest.
On comparing the observed flux modulation indices with the predictions, we find 9 of the measurements
to be in reasonable agreement with their predicted values.
For \elevb, the measured value (\mr = 0.21) is much below the prediction (0.35).
For rest of the data (9 measurements), the observed values are substantially larger than the predictions.
According to Romani et al. (1986), for a spectrum with $\alpha = 4$, the flux modulation index for an
extended medium is expected to be $\sqrt{3}$ times larger than that for a thin screen model.
This would mean \mr $\approx$ 0.66, much above any of the measured values.
Thus, excluding one measurement, the observed flux modulations are consistent with 
$\kolind \le \alpha < 4 $,
if one considers the scattering material to be uniformly distributed along the line of sight.

Our analysis shows that the modulations due to various sources of 
errors and non-ISS effects (Appendices A and B) can be completely ignored for \egtd and PSR B0919+06.
These data are characterized by reasonably good statistical reliability, both in terms of number of 
measurements and number of refractive cycles of fluctuations (Table 3).
The contributions to the modulation indices (of \nd, \td and F) due to the measurement noise are only 
marginal for these data.
Also, flux modulations due to Faraday rotation effects are negligible for \egtd ($ \mrpol \approx 0.01 $)
and marginal for PSR B0919+06 ($ \mrpol \approx 0.14 $).
Further, effects due to the Earth's orbital motion (\deltvobs) and the motion of the density irregularities 
(\deltvirr) can be ignored for PSR B0919+06 and are only marginal for \egtd (a worst case reduction of 
0.02 in \mt on accounting for both the effects).
In addition, the data are free from effects such as persistent drift slopes and systematic trends in the 
time series.
The results show that the modulation indices of all the 3 quantities are considerably above the Kolmogorov 
predictions.
The measurements of \mb are consistent with $ \kolind < \alpha \le 4 $ and that of \mt with 
$ \alpha \approx 4 $, whereas those of \mr require $ 4 \le \alpha < 4.3 $, as far as the predictions 
of a thin screen model are concerned.
If one considers an extended medium, there is agreement between the measured and predicted values 
of \mr for \egtd (0.33 versus 0.32), but for PSR B0919+06, the measured value is considerably larger than 
the prediction (0.46 versus 0.28).

In summary, the measured modulation indices of \nd, \td and F 
are found to be considerably larger than the predictions of a 
thin screen model with a Kolmogorov form of density spectrum (hypothesis IA of \S 1).
Effects due to various sources of noise involved in the measurements are insignificant except for a few 
values.
Further, modulations due to non-ISS effects such as variable Faraday rotation and the Earth's orbital 
motion can be ignored for most of the data.
If we compare the flux modulation indices with the predictions of a uniform scattering medium along 
the line-of-sight, roughly half of the measurements agree with the predicted values.
However, similar predictions are not available at present for the modulation indices of \nd and \td.
Clearly, theories based on a thin screen model and hypothesis (IA) 
are inadequate to account for the results from the present observations.

%%%%%%%%%%%%%%%%%%%%%%%%%%%%%%%%%%%%%%%%%%%%%%%%%%%%%%%%%%%%%%%%%%%%%%%%%%%%%%%%%%%%%%%%%%%

\subsection{Drifting of Intensity Patterns in Dynamic Spectra}

Another important observable effect of refractive scintillation which can be studied using our data is 
drifting bands in dynamic spectra.
The problem was first addressed by Shishov (1973), and subsequently by Hewish (1980).
The basic picture is that the diffraction by small-scale irregularities (typically $ 10^6 $ m to $ 10^8 $ m) 
results in an angular spectrum of scattered rays, and its mean direction of arrival is modified by the
refraction through large-scale irregularities (typically $ 10^{10} $ m to $ 10^{12} $ m).
The bending angle, \refr, usually known as `refractive scattering angle',
is determined by the phase gradient due to large-scale density irregularities, and given by

\begin{equation}
\refr ~ = ~ \left( { { \lambda \over 2 \ \pi } \nabla \phiref } \right) 
\propto \left( { \partial \nele \over \partial r } \right)
\end{equation}

\noindent
where $ \lambda $ is the observing wavelength, $r$ is the transverse dimension, 
and \phiref is the slowly varying phase component (sometimes known as ``refractive phase'') 
and \nele is the electron number density.
The intensity patterns at the observing plane are displaced by X $ \sim $ Z\refr, 
where Z is the distance to the phase screen.
The frequency dependence of the refraction angle (\refr $\propto$ $\nu ^{-2}$) results in varying
magnitudes of displacements for patterns at different frequencies.
In the presence of a relative motion between the pulsar and the observer, these cause intensity peaks 
at different frequencies to arrive at progressively increasing delays and hence appear as sloping patterns 
in the dynamic spectra.
On elaborating the analytical treatments given by Shishov (1973), Hewish (1980) showed that the drift
slope, \dnt, can be related to the refractive steering angle, $ \refr $, through the following expression

\begin{equation}
{ d t \over d \nu } ~ = ~ { {\rm D} \ \refr \over \viss \fobs }
\end{equation}

\noindent
where \viss is the speed of scintillation patterns, \fobs is the frequency of observation, and D is the
separation between the source and the observer.
The above expression is for a thin screen geometry, with the screen located midway
between the source and the observer (D=2Z). 
In characterizing the drifting features in our data, we prefer \dtn over \dnt.
Justification for this choice and our definition of drift slope is described in Paper I.

Drifting of intensity patterns are extensively seen in our data (e.g. Figs. 1(a)$-$(h) of Paper I). 
The property is highly pronounced for PSRs B0834+06, B1133+16, B1919+21 and B2045$-$16.
The measured values of \dtn 
(see Figs. 4(a)$-$(x) of Paper I) range from $ \sim $ 0.05 \dtnu (e.g. PSRs B1133+16, B1237+25) 
to a few \dtnu (e.g. PSRs B1604$-$00, B2327$-$20).
Also several pulsars show gradual and systematic variations of drift slopes, 
along with a number of sign reversals, during the time span of observations.
The data of PSRs B0823+26 and B0919+06 are good examples illustrating this property.
In general, drift slopes are found to vary over time scales comparable to those of \nd, \td and F,
but our data are not sampled regularly enough to obtain robust estimates of these time scales.
The drift slope averaged over all the measurements of a given data set, \avdtn, and the rms 
fluctuation, \rmsdtn, are computed for each data set, and are given in columns (3) and (4) of Table 7.
For several pulsars, values of \avdtn are found to be quite close to zero.
Some pulsars, especially those characterized by few or no slope reversals,
show significantly large values of average drift slopes.

%%%%%%%%%%%%%%%%%%%%%%%%%%%%%%%%%%%%%%%%%%%%%%%%%%%%%%%%%%%%%%%%%%%

\subsubsection{Pattern Drifts and Decorrelation Bandwidth}

Refractive effects, such as those which produce drifting intensity patterns, are expected to affect the 
estimation of the decorrelation bandwidth (e.g. Gupta et al. 1994; Cordes et al. 1986).
Usually, the decorrelation bandwidth is measured as the half-width at half-maximum 
along the frequency lag axis of the 2-D ACF of the dynamic spectrum.
In the absence of refractive bending, this method correctly estimates the true decorrelation 
bandwidth produced by DISS, as intensity patterns at different frequencies are aligned in time.
However, in the presence of significant refractive bending, this method will always underestimate the 
decorrelation bandwidth as the drifting patterns are no longer aligned in time.
Under such conditions, a better estimate can be obtained by measuring the half-power bandwidth along the
direction of the drift slope rather than along the frequency lag axis.
For the case where the direction of the shift of intensity patterns is aligned with the direction of the
scintillation velocity (\viss), the new technique will result in complete correction of the \nd value.
Clearly, the effectiveness of this technique decreases as the angle between the phase gradient and 
\viss increases from $0^o$ to $90^o$.

Thus we define a new estimator for the decorrelation bandwidth, which we refer to as the
``drift-corrected decorrelation bandwidth'', \ndc, which is the frequency lag corresponding to 
the point on the
half-maximum contour of the ACF that is furthest from the time lag axis.
In terms of the parameters \cone, \ctwo and \cthree describing the model Gaussian (eq. [2] of Paper I) 
fitted to the ACF, \ndc can be expressed as 

\begin{equation}
\ndc ~ = ~ {\rm \left( ln ~ 2 \right) ^{0.5} ~ \left( C_1 ~ - ~ { C_2^2 \over 4 ~ C_3 } \right) ^{-0.5} } .
\end{equation}

\noindent
The time series of corrected decorrelation bandwidths obtained in this manner are shown in Figs. 1(a)$-$(x) 
of this paper.
These can be compared with the `instantaneous' or `apparent' decorrelation bandwidth, \nd 
$-$ shown in Fig. 4 of Paper I $-$ to obtain some idea about the reduction in decorrelation bandwidth due to 
refractive effects.
We also compute the statistical properties of \ndc, such as its average value, \avndc, and fractional rms 
fluctuation, \mbc, which are given in columns (6) and (7) of Table 7.  Columns (8) and (9) of Table 7
give the noise modulation indices and the noise corrected values of \mbc, analogous to columns (3)
and (6) of Table 4.

Here we briefly summarize the general characteristics of the corrected bandwidth. 
For most pulsars, its average value is larger than that of the traditional
decorrelation bandwidth (i.e., $ \avndc > \avnd $).
In addition, modulation indices of \ndc are generally smaller than those of \nd (i.e., $ \mbc \la \mb $). 
Only exceptions are PSRs B1237+25 and B1604$-$00 for which estimates of \mbc are substantially larger 
than those of \mb. 
For PSR B1604$-$00, this is due to a few dominating measurements in the time series (Fig. 1.o).
For PSRs B1540$-$06 and B2310+42, we see that corrected bandwidths appear to be more or less 
stable (\mbc \la 0.1),
which could be an artifact due to the limitation of our spectrometer resolution.
Excluding these 4 outliers, we find the global average modulation index (\avmbc) to be  $\approx 0.3$.
Measurement noise is expected to cause a modulation index of 0.1 (see column (8) of Table 7), and hence its 
contribution to the measured value is only marginal.
The only case where noise modulations are significant is \ninea, where $ \mbc \sim \mnoise $ and we get 
$ \mbcriss \approx 0.06 $.
On comparing the estimates of \mbc with the theoretical predictions, we find them to be considerably 
larger than the Kolmogorov predictions.
However, in contrast to \mb, most measurements range between the predictions of 
$ \alpha = \kolind $ and $ \alpha = 4 $.

The above analysis confirms that the traditional estimator for decorrelation bandwidth (\nd) is biased due to
the presence of refractive drifts which are produced by the large-scale irregularities in the ISM.
The new estimator for the decorrelation bandwidth (\ndc) is less prone to bias due to such refractive effects.
Thus, \ndc should be a better choice for estimating effects due to purely diffractive scintillation
phenomena, $ie.,$ effects produced by the small-scale irregularities in the ISM.
In the following section, where we attempt to estimate effects of small-scale and large-scale
irregularities independently, we use \ndc as the estimator for the decorrelation bandwidth.

%%%%%%%%%%%%%%%%%%%%%%%%%%%%%%%%%%%%%%%%%%%%%%%%%%%%%%%%%%%%%%%%%%%%%%%%%%%%%%%%%%%%%%%%%%%%%%%%%%%%%

\subsubsection{Estimation of Diffractive and Refractive Scattering Angles}

Diffractive and refractive scattering angles are two useful indicators of the magnitude of electron density 
fluctuations at small ($ \sim $ $ 10^6 $ m to $ 10^8 $ m) and large ($ \sim $ $ 10^{10} $ m to $ 10^{12} $ m) 
spatial scales respectively.
From our data, we estimate these two angles, 
denoted as \diff and \refr respectively, at each epoch of observation.
The diffractive scattering angle at $ i^{th} $ epoch, \diffi, is given by

\begin{equation}
\diffi ~ = ~ \left( { c \over \pi ~ {\rm D} ~ \ndci } \right) ^{0.5}
\end{equation}

\noindent
where \ndci denotes the measurement at $i^{th}$ epoch of observation and $c$ is the speed of light.
We use the average diffractive angle (\avdiff) over the entire time span of observation to
characterize density fluctuations at small spatial scales.
Our values of \avdiff are given in column (5) of Table 8.

The refractive angle at $ i^{th} $ epoch of observation, \refri, is obtained from the estimates of 
drift rate, $(\dtn)_i$, and scintillation pattern speed, \vissi, at that epoch, using the relation

\begin{equation}
\refri = \left( { \vissi \fobs \over {\rm D } } \right) \left( { d t \over d \nu } \right) _i \qquad .
\end{equation}

\noindent
We note that when the gradient of the refractive wedge is not aligned with the pattern velocity,
the true refractive angle is given by $ {\rm \refri sec \psi _i } $, where $ \psi $ is the 
angle between the gradient and the velocity.
For simplicity, we assume $ \psi $ = 0 in estimating \refri, but address this issue later in
this section.
The pattern speed (\viss) is estimated from the measurements of decorrelation bandwidth and scintillation 
time scale, using the following expression

\begin{equation}
{\rm
\vissi ~ = ~ A_V ~ 
\left( D_{[kpc]} ~ \ndci _{[MHz]} \right) ^{0.5} ~ \left( \fobsGHz ~ \tdi _{[sec]} \right) ^{-1}
\hspace{1.0cm} \velu
}
\end{equation}

\noindent
where \tdi denotes the measurements at $ i^{th} $ epoch of observation, 
and we adopt $ {\rm A_V = 3.85 \times 10^4 } $ given
by Gupta et al. (1994).
Our measurements of refractive scattering angles obtained in the above manner are presented in Figs.
2(a)$-$2(x), in the form of the time series for each pulsar.

According to the models which treat refraction effects as random phenomena due to the low wavenumber
part of the underlying density power spectrum, \refr is expected to vary randomly about a zero mean value 
over refractive time scales (Rickett 1990; Rickett et al. 1984, Romani et al. 1986).
But treatments which consider refraction by a separate large-scale component 
(e.g. Shishov 1973; Hewish 1980) may allow non-zero mean values for \refr. 
The mean refractive angles (\avref) computed from the time series in Fig. 2 are listed in column (3) 
of Table 8.
For a number of pulsars, \avref $ \approx $ 0 within the measurement uncertainties 
(e.g. PSRs B0919+06 and B0823+26), while for some, \avref is found to be significantly 
different from zero; best examples for which are PSRs B0834+06 and B1919+21.
PSRs B1133+16 (data from sessions II and III), B1237+25, B1604$-$00, 
B1929+10 and B2045$-$16 also show statistically significant non-zero values for \avref.
For PSRs B1237+25 and B1929+10, such an effect may be attributed to the poor statistics 
in terms of limited number of measurements and/or dominance of a few measurements over the 
rest (see Figs. 2.l and 2.s).
PSRs B1133+16 and B2045$-$16 show non-zero \avref despite sufficiently good statistics 
(\nep = 25$-$35).
A closer inspection of their time series of \refr (Figs. 2.j, 2.k and 2.v) reveal that, though not highly 
pronounced, there are some signatures of persistent drifts lasting over typically several weeks.
A similar trend can also be seen for PSR B1604$-$00 (Fig. 2.o), 
but here the sampling is rather coarse.
Thus not all our data are in support of the expectations of models based on random refraction.

As mentioned earlier, in the case of two-dimensional refraction, the observer will measure a refractive
angle ${\rm  \refr = \Refr cos \psi }$, where $ \Refr $ is the `true refractive angle' and $ \psi $ is 
the `alignment angle'. 
This effect can potentially modify the statistical properties estimated for \refr.
In particular, it will result in an underestimation of the rms refractive angle, \refrms. 
Though an exact correction for this is not practical, we attempt a first order correction by assuming 
$ \Refr $ and $ \psi $ to be independent zero-mean random variables, and also that $\psi$ can range from 
$-\pi/2$ to $\pi/2$ with uniform probability.
This gives $ \varRef = 2 \ \varref $, which implies $ \Rmsref = \sqrt {2} \ \rmsref $. 
This will yield somewhat better estimates of \rmsref, at least for data with \refmean $\approx$ 0. 
The values of \rmsref given in column (4) of Table 8 are the measured values scaled by $\sqrt{2}$.

%%%%%%%%%%%%%%%%%%%%%%%%%%%%%%%%%%%%%%%%%%%%%%%%%%%%%%%%%%%%%%%%%%%%%%%%%%%%%%%%%%%%%%%%%%%%%%%%%%%%%%%%%%%

\subsubsection{Estimation of Slope of the Electron Density Spectrum}

In this section, we use the measurements of \diff and \refr to estimate the slope ($ \alpha $) needed
for representing the underlying density fluctuations by a simple power-law form of spectrum, at least 
over the DISS and RISS scales of interest for our data. 
There are two possible ways of doing this.
The first is to make use of the ratio \rmsref/\avdiff as a discriminator of $\alpha$.
For $\alpha < 4$, the refractive scattering angles are expected to be smaller than the diffractive angles
($\rmsref < \avdiff$), but for `steep' spectra ($\alpha > 4$), $\rmsref > \avdiff$ can be expected (Hewish 
et al. 1985; Romani et al. 1986; Rickett 1990).
Earlier studies (Smith \& Wright 1985; Hewish et al. 1985) did not have good enough data to facilitate 
accurate determinations of \rmsref and \avdiff, and often employed \refr/\diff from a given (or a few) 
epoch(s) of observations to discriminate between different kinds of density spectra.
For the values of \rmsref and \avdiff given in Table 8, 
we find the ratio for most of the pulsars to be in the range 0.1$-$0.8, and
there is no pulsar for which it is above unity.
Thus, if the density spectra are to be represented by simple power-law forms, then the measurements 
of \diff and \refr from our observations will preclude $ \alpha > 4 $ for these epochs.

The second method is to estimate the power levels \pref and \pdiff at refractive and diffractive
wavenumbers \kref and \kdiff respectively, from which the slope of the density spectrum can be 
estimated (e.g. Armstrong et al. 1995). 
The procedure is as follows.
From the measurements of \rmsref, one can obtain the structure function level at refractive
length scale \sref (given by $1/\kref$), from which an estimate of the amplitude of the density 
spectrum (\cN) is obtained. 
The power level at refractive scale is then given by \pref = \cN \krefalp.
From the measurements of decorrelation bandwidths, one can estimate the amplitude of the spectrum (\cn) at
small spatial scales, and the power level at diffractive wavenumber (given by $\kdiff=1/\so$, where \so 
is the `coherence scale') is then given by \pdiff = \cn \kdiffalp.
This method should give $\cN \approx \cn$ if the assumed value of $\alpha$ is correct. 
Alternatively, one can estimate a slope $\beta$ = log (\pref/\pdiff)/log (\kref/\kdiff).
Throughout this paper, we use $\alpha$ to denote the power-law index (as defined in eq. [1]) of the density
spectrum, and $\beta$ to represent the slope estimated from our measurements.

It is easy to show that the above two methods are not independent and also that for $\alpha < 4$, 
there is an exact correspondence between the ratio of scattering angles (\rmsref/\avdiff) and the 
slope estimate, $\beta$. 
For this, we first rewrite equation (3) of Armstrong et al. (1995) as

\begin{equation}
\cng ~ = ~ K_{\alpha} ~ \left[ \Dphi \left( s \right) \right]
~ \left( 8 ~ \pi ~ r_e^2 ~ \wavobstwo ~ {\rm D } ~ s ^{\ald} \right) ^{-1}
\end{equation}

\noindent
where $\ald = \alpha - 2$, 
$ K_{\alpha} = \left( 1 + \ald \right) \left[ f \left( \ald \right) \right] ^{-1} $,
D is the propagation distance, and $r_e$ is the classical electron radius. 
The phase structure function, \Dphi, at refractive scale, \sref, is given by

\begin{equation}
\Dphi(\sref) ~ \approx ~ \left[ \left( { 2 \pi \over \wavobs } \right) \rmsref \sref \right] ^2
\end{equation}

\noindent
where \wavobs is the observing wavelength. 
The refractive scale can be taken as the `scattering disk', given by $ \sim $ D\avdiff.
The structure function is unity at coherence scale (\Dphi(\so) = 1), which is also considered as the 
diffractive length scale, \sdiff.
Thus, from equation (13), we can get the two estimates of the amplitude of the density spectrum $-$
\cN and \cn. 
Using these expressions, the slope estimate, $\beta$, simplifies to 

\begin{equation}
\beta = 4 + \left( { {\rm log} \left( { \rmsref / \avdiff } \right) \over {\rm log} \left( u \right) } \right)
\end{equation}

\noindent
where $u$ is a measure of the the strength of scattering, defined as  $\sqrt{\sref/ s_o}$ (Rickett 1990).
The value of $u$ can be estimated from decorrelation bandwidth measurements (see Table 4 Paper I for our estimates).
We note that equation (15) is very similar to equation (2.9) given by Rickett (1990).
We also note that the above result is valid only for $\alpha < 4$, as equation (13) is valid only for this
regime.
From this result, one can see that, for a given value of $u$, the ratio of scattering 
angles is related to the slope of the density spectrum.
Further, one can see that as $\beta \longrightarrow 4$, $\rmsref \longrightarrow \avdiff$.

Our estimates of $\beta$ are given in column (6) of Table 8, and they range from 3.3 to 3.9.
Since all the 3 quantities (\rmsref, \avdiff and $u$) have been fairly accurately determined, 
we are able to estimate the $\beta$ values with fairly
good accuracies (1-$\sigma$ uncertainty $ \sim $ 0.02$-$0.1).
In some cases, the uncertainties are larger ($ \sim $ 0.1$-$0.2), probably due to sloping patterns being 
not well pronounced in the corresponding data.
Further, none of our $\beta$ values are above the critical value 4, which gives an ex post facto 
justification of the assumption $\alpha < 4$ made in obtaining equation (15).

Taking into consideration the uncertainties (at $\pm$2-$\sigma$ levels), 
we find 18 of the 25 measurements to be consistent with the Kolmogorov value 11/3. 
For 6 data sets, the measured values are significantly above 11/3.
The highest value ($\beta = 3.91 \pm 0.03$) is measured for PSR B1929+10, the closest pulsar in our sample.
The other cases are PSRs B1604$-$00 and B2327$-$20, and part of the data of PSRs B0834+06 (session II)
and B1133+16 (sessions II and III), with $\beta \approx$ 3.77 to 3.8.
We note that values of $\beta$ that are significantly below \kolind ($ie, ~ \beta \approx 3.3 - 3.4 $)
are also the ones that have large measurement errors (PSRs B0329+54, B1508+55, B1540$-$06 and
B2310+42).
Further, these are also the cases where sloping patterns are less pronounced, due to which one tends to
underestimate the \dtn values and consequently the $\beta$ values.
For one data set (\egta), the estimated $\beta$ ($\approx 3.58 \pm 0.03$) is significantly below the 
Kolmogorov index. 
Thus, not all the measurements from our data are consistent with a Kolmogorov form of density spectrum
(hypothesis IA of \S 1).
The larger values of $\beta$ (3.77 to 3.91) are seen for nearby pulsars ($ \sim $ 200 to 700 pc) and there 
is a weak trend for a decrease in $\beta$ with DM (up to $\sim$ 20 \dmu) and 
distance (up to $\sim $ 1 kpc) (see Figs 3.a and 3.b).
Clearly, a ubiquitous nature of the density spectrum is not quite supported and there are
several directions towards which the density spectrum appears to be steeper than $\alpha = 11/3$.
 
Before proceeding further, we comment on some of the underlying assumptions in our $\beta$ estimation. 
Firstly, our $\beta$ values are indicative of the true slope only if (i) the spectrum is a simple power-law, 
and (ii) $ \alpha < 4 $. Secondly, the method is subject to the validity of the implicit assumption of 
drifting features arising due to density fluctuations on spatial scales $ \sim $ scattering disk, D\diff. 
Further, the method assumes a stationary statistics for RISS, which need not be necessarily true in
practice, especially for data with non-zero \avref. Hence \rmsref, and consequently $\beta$, should be 
treated with caution for those cases. In the simplest model of a thin screen placed between the source and 
the observer, it is easy to see that the method is insensitive to the location of the screen.

%%%%%%%%%%%%%%%%%%%%%%%%%%%%%%%%%%%%%%%%%%%%%%%%%%%%%%%%%%%%%%%%%%%%%%%%%%%%%%%%%%%%%%%%%%%%%%%%%%%%%%%%%%%

\subsection{Persistent Drifting Features in Dynamic Spectra}

According to the theoretical models for RISS which consider the underlying density fluctuations to be a 
stochastic process, drift slopes of patterns are expected to vary randomly about a zero mean value over 
refractive time scales (cf. Rickett 1990).
From the time series of measurements (see Fig. 4(a)$-$(x) of Paper I), we see that this broad picture is 
substantiated by a number of pulsars, of which PSRs B0823+26 and B0919+06 form good examples.
However, as mentioned earlier, our data also show several examples where the drifting features are 
`persistent' and do not show frequent sign reversals of slopes.
A visual examination reveals that data from the first 3 observing sessions of PSR B0834+06 and from the 2 
sessions of PSR B1919+21 form the best examples of such persistent slopes, as they are characterized by a 
complete absence or few epochs of slope reversals (e.g. \egta, \nineb). 
A closer inspection of Fig. 4 of Paper I (and also Fig. 2 of this paper) reveals that
there are several other pulsars which form comparatively weaker examples of such a property;
PSRs B0329+54, B0823+26(II), B1540$-$06, B2020+28 and B2310+42 belong to this category.

To make a quantitative distinction between the two cases, $viz.$ 
``frequent drift reversals'' and ``persistent drift slopes'',
we examine the distributions of the measured drift slopes for a clear ``skewness'' with respect to zero.
Selected plots of such distributions are shown in Figs. 4(a)$-$(d) to illustrate this scheme.
The idea here is to identify data for which the mean drift slope, \avdtn, is substantially offset from 
zero in comparison to its observed fluctuations.
Data for which mean slopes are smaller than the rms fluctuation (\avdtn $ < $ \rmsdtn) 
and are free from a skewness in the distribution (e.g. Fig. 4.b) are categorized as 
Class I, and those with mean slopes offset from zero by more than the rms (\avdtn \ga \rmsdtn) 
and have a skewed distribution (e.g. Figs. 4.a and 4.c) are labeled as Class II.
The classification becomes ambiguous when the quantities \avdtn and \rmsdtn have substantial uncertainties,
and the skewness is not well pronounced; these are treated as `NC' (non-classifiable).
Column (5) of Table 7 shows the `drift class' decided in this manner.

As per the above scheme, 
PSRs B0834+06(I), B0834+06(II) and B0834+06(III), B1919+21(I) and B1919+21(II) come under Class II. 
For PSRs B0329+54, B0823+26(II), B1133+16(III), B1237+25, B2045$-$16 and B2310+42, 
the estimates of \avdtn and \rmsdtn are comparable, but the uncertainties are large;
hence, we treat them as NC.
Rest of the data are free from such ambiguities and belong to Class I.
A special pulsar in this context is PSR B0834+06, which shows the behaviour of Class I 
in the final observing session, much in contrast with what is seen in the first three sessions.

Earlier studies of dynamic spectra (Gupta et al. 1994) reported a similar property for 
PSRs B0628$-$28 and B1642$-$03.
For PSR B0628$-$28, no persistent drifting features are seen in our data.
As mentioned before, our observations show PSR B0834+06 changing the behaviour from persistent drift slopes 
(January 1993$-$June 1994) to frequent slope reversals (April$-$July 1995). 
For PSR B1919+21, we do not have similar information, as it was not followed-up for a third session.
From these examples, it appears that persistent drifting features usually last over time intervals 
$ \sim $ several months to a few years.

%%%%%%%%%%%%%%%%%%%%%%%%%%%%%%%%%%%%%%%%%%%%%%%%%%%%%%%%%%%%%%%%%%%%%%%%%%%%%%%%%%%%%%

\subsubsection{Implications for the Density Spectrum}
 
We now examine the implications of our observations of persistent drift slopes for the nature of 
the density irregularity spectrum.  
The crucial question is whether the density structures responsible for such effects form part of 
the power-law spectrum determined by measurements of \rmsref and \avdiff, as described in \S 2.2.3.
This can be answered by extrapolating the power-law spectrum to $ \kappa = 1 / S $
(where $S$ corresponds to the time span over which persistent drift slopes are observed),
estimating the expected rms refractive angle that would be produced by the
power at these scales, and comparing this with the measured values of \avref.
The expected rms refractive angle can be obtained simply by inverting equation (15), while using
$U = \sqrt{S/s_o}$ in place of $u$.  Since \avbeta $\approx$ 11/3 for both PSRs B0834+06 and B1919+21,
this gives $\refkol = \avdiff \ U ^ {-1/3}$.
These values are given in column (7) of Table 9.
Since the measured values of \avref for individual sessions are $\sim$ 1.2$-$2.7 times larger than the \refkol
values, the probability of the density structures being part of a Kolmogorov-like spectrum is {\it 
rather low} ($\sim$ 1$-$3\% for PSR B0834+06 and $\sim$ 15$-$20\% for PSR B1919+21).
Using the larger values of $S$ for the combined data from multiple sessions increases the discrepancy with
the measured \refkol (\avref is 3.4 times larger for PSR B0834+06 and 1.8 times for PSR B1919+21) and
further reduces the probability of the structures being part of the power-law spectrum determined by 
\rmsref and \avdiff (0.1\% for PSR B0834+06 and 6\% for PSR B1919+21).
  
In Figs. 5.a and 5.b, we have plotted the power levels at wavenumbers corresponding to diffractive 
and refractive scales (\sdiff and \sref), and at the larger spatial scales ($S$).
The power levels at $ \kappa = 1 / S $ are significantly above the Kolmogorov expectations,
which can be interpreted in different ways.
One possibility is that a single power-law description is inadequate and the spectrum steepens at lower
wavenumbers ($ 10^{-14} \ {\rm m^{-1}} \ \la \ \kappa \ \la \ 10^{-11} \ {\rm m^{-1}}$).
This would correspond to the type IIIA spectrum of \S 1.
From the estimated power levels, we find the average slope over this range (\betasteep) to be much larger 
than \kolind ($\approx$ 4.9 for PSR B0834+06 and $\approx$ 4.5 for PSR B1919+21).
Such a ``piece-wise power-law'' form of spectrum for representing the density fluctuations over a wide 
range of spatial scales (about 6 decades or more) is an interesting possibility.
The other possibility is that a Kolmogorov-like distribution of irregularities is superposed with 
a separate large-scale component giving rise to a ``bump'' near 
$ \kappa \ \sim \ 10^{-13} - 10^{-12} \ {\rm m^{-1}} $ (type IIIB spectrum of \S 1).
With the present observational data, it is difficult to discriminate between these two options.

%%%%%%%%%%%%%%%%%%%%%%%%%%%%%%%%%%%%%%%%%%%%%%%%%%%%%%%%%%%%%%%%%%%%%%%%%%%%%%%%%%%%%%

\subsubsection{Constraints on Discrete Plasma Structures}

A possible alternative interpretation of the persistent drifting features is existence of 
large-scale deterministic density structures along the line-of-sight to the pulsar.
Our observations allow us to put constraints on the characteristics of such structures, 
in particular their sizes and electron densities.
If we consider a refractive wedge with thickness L and electron density \Nele, the resulting refractive angle 
\Refr is given by 

\begin{equation}
\Refr ~ = ~ { 1 \over k } ~ \left( { \partial \phi \over \partial r } \right) ~ = ~
\left( { r_e ~ \wavobs \over k } \right) ~ \int _0 ^L { \partial \Nele \over \partial r } dr
\end{equation}

\noindent
Under the assumption that \Nele is uniform within the wedge, the integral simplifies to 
$ {\rm \Delta \Nele L } / S $,
where $\Delta \Nele$ is the deviation from the ambient mean density 
($ie.,~\Delta \Nele = \Nele - \avne$)
and $S$ is the transverse extent of the wedge.
In the simplest case of a spherical cloud, the `aspect ratio' is unity ($S$ $ \sim $ L), and assuming high 
electron densities ($ \Nele \gg \avne $), the above expression simplifies to 

\begin{equation}
\Refr ~ = ~ { r_e ~ \wavobstwo ~ \Nele \over 2 ~ \pi }
\end{equation}

\noindent
where $ r_e $ is the classical electron radius ($ 2.82 \times 10^{-15} $ m) and \wavobs is the 
observing wavelength.
We essentially estimate the electron density (\Nele) required to produce the mean refractive angle, \avref.  
The constraint on the size ($S$) is simply given by

\begin{equation}
S ~ \ga ~ \avvissc \tsp
\end{equation}

\noindent
where \tsp is the time span of observation over which drifting features lasted (column 4 of Table 9), 
and \vissc ($ie,$ \viss computed using \ndc and \td) is given by equation (12) (column 3 of Table 9).
The inferred values of \Nele and $S$ are estimated for PSRs B0834+06 and B1919+21, 
and are listed in columns (5) and (6) of Table 9.

The mean refractive angles required to produce the persistent drifting features seen in our data are moderate 
(0.1 to 0.3 mas).
The important implication is that high electron densities (\Nele $ \sim $ 2$-$4 \cmc) need to persist over 
spatial scales much larger than the characteristic refractive scales ($S$ $ \gg $ \sref) in order to give 
rise to such effects.
Constraints on the size of these structures from a single observing session (\tsp $\sim$ 100 days) 
are $S$ $\sim$ 10 AU.  Further, both these pulsars show similar persistent drifting features for more
than one session.
If we assume that the persistent drifts are sustained during the intervals between the successive sessions, 
then \tsp is much longer ($\sim 300 - 500$ days), and the inferred sizes are $\sim$ 70 AU for PSR B0834+06
and $\sim$ 40 AU for PSR B1919+21 (see Table 9).

%%%%%%%%%%%%%%%%%%%%%%%%%%%%%%%%%%%%%%%%%%%%%%%%%%%%%%%%%%%%%%%%%%%%%%%%%%%%%%%%%%%%%%%%%%%%%%%%%%%%%%%

\section{Discussion}

We have studied the properties of DISS and RISS for a number of nearby pulsars in an attempt to 
constrain the power spectrum of plasma density fluctuations in the ISM. 
We have focused on the results from two important and easily observable effects
due to refractive scintillation: (i) modulations of DISS observables (\nd and \td) and flux density,
and (ii) drifting bands in dynamic spectra.
Our sample consists of mostly nearby pulsars (D $ \la $ 1 kpc), and
there is a reasonably uniform coverage in $(l,b)$, DM and D; hence, a more or less unbiased sample.
Our data are sensitive to the density inhomogeneities in the spatial scale range $\sim$ $10^7 - 10^{13}$ m. 
Since all the basic measurements of DISS (\nd and \td) and RISS (\dtn, \mb, \mt and \mr)
used in our analysis are from self-consistent data sets, the possibility of an observational bias is reduced.
Furthermore, we have relied upon the more meaningful quantity, \rmsref/\avdiff, for discriminating between 
different kinds of density spectra and estimating $ \beta $ values,
compared to the earlier attempts which often employed estimates of \refr/\diff from one
(or a few) epoch(s) of observations.
As discussed in Paper I, it has also become possible from our observations to estimate the average
scintillation properties more robustly than previously published work.
Therefore, we believe the implications of our results for the nature of the electron density spectrum need
serious consideration.

%%%%%%%%%%%%%%%%%%%%%%%%%%%%%%%%%%%%%%%%%%%%%%%%%%%%%%%%%%%%%%%%%%%%%%%%%%%%%%%%%%%%%%%%%%%%%%%

\subsection{Implications of the Main Results for the Density Spectrum}

The main results from our data can be summarized as follows:

\begin{enumerate}
\item
Our observations show large-amplitude modulations of decorrelation bandwidth (\nd), scintillation time scale 
(\td) and flux density (F). The measured depths of modulations are found to be considerably larger than the 
predictions of a thin screen model with a simple Kolmogorov form of density spectrum (hypothesis IA of \S 1).
Barring some cases, the measured modulation indices of \nd and F are consistent with $ 4 < \alpha < 4.3 $,
and those of \td with $ \kolind < \alpha < 4.3 $, as far as the predictions of a thin screen model are 
concerned. For the flux modulation indices, better agreement is seen with spectrum of type IA 
if the scattering medium is taken to be uniformly distributed along the line of sight. 
Even then, roughly half the measurements are significantly larger than the predictions of type IA spectrum.
\item
Measurements of refractive and diffractive angles are consistent with $ \alpha < 4 $,
as the ratio \rmsref/\avdiff is found to be below unity for all pulsars.
Further, our estimates of density spectral slope ($\beta$) range from 3.3 to 3.9.
While 18 of the 25 measurements are consistent with the Kolmogorov index (at $\pm$2-$\sigma$ levels), 
for 6 pulsars (D $ \sim $ 200 to 700 pc), $\beta$ is found to be significantly larger than 11/3.
\item
Persistent drifting bands lasting over many months are seen with PSRs B0834+06 and B1919+21, which imply
excess power at spatial scales $ \sim $ 10$-$100 AU (much larger than the refractive scales) compared to 
the expectations from a type IA spectrum. 
This is possible if the spectrum (i) is piece-wise power-law which steepens at 
$ \kappa < \kref $ or (ii) has a low wavenumber enhancement (hypotheses IIIA and IIIB respectively of \S 1).
An alternative possibility is the existence of localized density structures of spatial scales $\sim$ 10$-$70 
AU and \Nele $\sim$ 2$-$4 \cmc (hypothesis IV of \S 1). 
\end{enumerate}

%%%%%%%%%%%%%%%%%%%%%%%%%%%%%%%%%%%%%%%%%%%%%%%%%%%%%%%%%%%%%%%%%%%%%%%%%%%%%%%%%%%%%%

\subsubsection{Refractive Modulations and Slopes of the Density Spectrum}

It is difficult to reconcile the first two results with a simple power-law model for the electron density
spectrum in the ISM.
Hence we look at them more critically.

Starting with the first result, we note that ours are not the first reported measurements of modulation 
indices larger than the Kolmogorov expectations.
Flux monitoring observations have been made by a number of groups earlier 
(Stinebring \& Condon 1990; Kaspi \& Stinebring 1992; Gupta et al. 1993; LaBrecque et al. 1994; Gupta et
al. 1994).
In Table 10, we summarize the results from all these observations, for pulsars common with our
observations.
To compare the measurements made at different frequencies, we use equation (4) for scaling the predicted
modulation indices.
We find that our measured flux modulation indices (column (3)) are comparable with observations at
nearby frequencies (e.g. columns (4), (5), (8) and (9)).
Further, most of the modulation indices given in Table 10 are significantly larger than those expected from a
Kolmogorov spectrum (type IA of \S 1).
For modulation indices of \nd and \td, there are very few observations reported in the literature that we
are aware of.
Gupta et al. (1994) have reported \nd modulation indices for 6 pulsars from their long-term scintillation
study at 408 MHz (their Table 4).
The values for the 5 pulsars common with our observations are comparable, and more than the predicted
values for type IA spectrum.

As discussed in \S 2.1.3, if the scattering medium is assumed to be uniformly distributed along the 
line of sight, our results of flux modulation indices are in somewhat better agreement with a type IA
spectrum.
About half the values are then consistent with $\alpha = \kolind$ and the remaining half are consistent
with $\kolind < \alpha < 4$. 
Due to the lack of relevant predictions,
similar comparisons cannot be made for the modulations of \nd and \td.
Thus our results from modulation indices can be partially reconciled with our $\beta$ estimates.

Turning now to the second result, we note that it may be possible that our refractive angle measurements 
(and hence the $\beta$ values derived from them)
underestimate their true values, despite the first order correction applied 
for the effect due to the alignment angle $\psi$  (\Rmsref = $\sqrt{2}$ \rmsref).
The underlying assumption of random fluctuations of \Refr and $\psi$ need not be true always, 
in which case the above correction may be still inadequate.
Observations of persistent drifts (for 2 pulsars) and statistically significant non-zero values of 
\avref (for 5 pulsars) in our data indicate that such situations exist in practice.
In such cases, the true $\beta$ values can be larger than our estimates, thereby reducing the disagreement
with the first result.

Another aspect worth mentioning is the assumption of isotropic turbulence in the ISM.
It is known that the presence of a strong magnetic field will make the turbulence highly anisotropic
(Higdon 1984, 1986), a concept well supported by the recent observations of field-aligned anisotropic
density structures in the inner solar wind (Armstrong et al. 1990; Anantharamaiah, Gothoskar \& Cornwell 1994).
However, it is unclear at present what is the importance of anisotropy in the ISM. 
As Coles et al. (1987) point out, there may be several important consequences for ISS in the case of 
anisotropic turbulence, but they have not been worked out analytically. 
The anisotropic turbulence will make \cn sensitive to the field geometry and can also affect the inner scale
cutoff; hence, it can potentially influence the modulations of DISS observables and flux.
Since the fluctuations in the large-scale Galactic magnetic fields have length scales $ \sim $ 100 pc 
(Simonetti, Cordes \& Spangler 1984), anisotropic ISS is probably more relevant for nearby pulsars.
Furthermore, there is evidence for large-scale ($ \sim $ 100$-$500 pc) spatial inhomogeneities in \cn 
within the local (\la 1 kpc) ISM (Bhat et al. 1997, 1998). 
Therefore, anisotropic ISS may be relevant for some of our pulsars.

%%%%%%%%%%%%%%%%%%%%%%%%%%%%%%%%%%%%%%%%%%%%%%%%%%%%%%%%%%%%%%%%%%%%%%%%%%%%%%%%%%%%%%%%%%%%%%%%

\subsubsection{Persistent Drift Slopes and Multiple Imaging Events}

In addition to persistent drifting bands, phenomena such as multiple imaging events 
(e.g. Wolszczan \& Cordes 1987) and extreme scattering events (e.g. Fiedler et al. 1987)
are also thought to be caused by large-scale deterministic density structures in the ISM.
While ESEs are mostly osberved with compact extra-galactic radio (EGR) sources, the other 
two effects are seen in pulsar dynamic spectra. 
So far, multiple imaging events have been reported for 7 pulsars $-$
Hewish et al. (1985) for PSRs B1133+16 and B1642$-$03;
Cordes \& Wolszczan (1986) for PSRs B0919+06, B1133+16 and B1919+21;
Wolszczan \& Cordes (1987) for PSR B1237+25; 
Kuz'min (1992) for PSR B1919+21;
Gupta et al. (1994) for PSR B2016+28; 
Rickett, Lyne \& Gupta (1997) for PSR B0834+06;
Bhat et al. (1998a) for PSR B1133+16.
Very few observations of persistent drifts have been reported so far.
For PSRs B0834+06 and B1919+21, our data show persistent drift slopes lasting over $ \sim $ 300$-$500 days.
A similar property is reported by Gupta et al. (1994) for PSRs B0628$-$28 and B1642$-$03, 
and Smith \& Wright (1985) see some signatures of it for PSRs B0823+26 and B1929+10. 
PSR B1937+21 is the only pulsar for which the occurrence of ESEs has been reported. 
To date, a total of 10 ESEs have been identified with EGR sources (see Fiedler et al. 1994 for a summary).

Fiedler et al. (1987) infer a very high density cloud (\Nele $\sim$ 1000 \cmc) of $\sim$ 7 AU from the
observations of ESEs in the light curves (at 2.7 and 8.1 GHz) of quasar 0954+658.
This unusually large constraint on the density can, however, be relaxed to \Nele $\sim$ 100 \cmc if one
considers an edge-on geometry with an aspect ratio ($\eta$) of 10:1.
Cognard et al. (1993) infer clouds of similar densities ( \Nele $\sim$ 25$-$220 \cmc) from the ESE
observed for PSR B1937+21 at 1.4 GHz.
However, the requirement on the size is much smaller ($\sim$ 0.05$-$0.1 AU) as the event spans only 
a short period of 15 days.
Rickett et al. (1997) suggest a structure of $\sim$ 3 AU and \Nele $\sim$ 40 \cmc (for $\eta = 1$) 
as one of the possible scenarios to explain the interstellar fringes observed at 408 MHz for PSR B0834+06.
Compared to all these results, the structures inferred from our data are comparatively larger in size,
but relatively less dense.

The similarity in the interpretations of the above three effects indicates a probable connection between them.
Romani, Blandford \& Cordes (1987) point out that the periodicities in pulsar dynamic spectra, LFV of quasars
and the ESE of 0954+658 (Fiedler et al. 1987) can be understood in terms of multiple imaging and focusing
by large-scale refracting structures in the ISM.
It is interesting to note that PSRs B0834+06 and B1919+21, which show persistent drifts in our observations, 
are known to have shown multiple imaging events 
(Cordes \& Wolszczan 1986; Kuzm'in 1992; Rickett et al. 1997).  
Similarly, PSR 1133+16, which shows evidence for multiple imaging in our data (also see Hewish et al. 1985; 
Cordes \& Wolszczan 1986), also shows some signatures of persistent drifts lasting over several weeks (see 
Figs. 2.j and 2.k).
The large proper motion of this pulsar ($\approx $ 475 \velu) would mean the corresponding spatial scales 
to be $ \sim $ 10 AU, quite comparable to the size of the density structures inferred from the persistent 
drifts of PSRs B0834+06 and B1919+21 (see Table 9). 
Thus, our observations of these three pulsars along with detections of multiple imaging events in earlier
observations form a direct evidence in favour of the connection between the two effects, thereby supporting 
the view of Romani et al. (1987) on scattering effects due to localized high-density structures. 
It is not clear whether there is a large population of such structures in the Galaxy, 
but the observational data suggest hypothesis (IV) is relevant at least for some lines of sight.

%%%%%%%%%%%%%%%%%%%%%%%%%%%%%%%%%%%%%%%%%%%%%%%%%%%%%%%%%%%%%%%%%%%%%%%%%%%%%%%%%%%%%%%%%%%%%%%%

\subsection{A Summary of Various Constraints on the Plasma Density Spectrum}

A number of attempts have been made in the recent past towards determining the form of the density
spectrum, and there are conflicting interpretations from various kinds of measurements.
While several observations are consistent with a simple Kolmogorov form (hypothesis IA), there is 
substantial amount of observational data which go against it.
Attempts have also been made to construct a composite spectrum extending over a wide range of spatial scales.
Here we give an overview of various observational evidences accumulated so far $-$ from our data as well as
from the published literature.

Among the several possible alternatives to the {\it pure} Kolmogorov form (hypothesis IA), three specific 
cases $viz,$ (i) steeper spectra (hypothesis II), (ii) Kolmogorov spectrum truncated at a large inner scale
(hypothesis IB), and (iii) Kolmogorov spectrum with a low wavenumber enhancement (hypothesis IIIB) have been 
more commonly discussed in the literature. 
Effects due to such {\it non-Kolmogorov} forms of spectra are not fully understood yet. 
Something that is common to all the three is that they are of more refractive nature than 
case IA. 
Most observational evidences against hypothesis IA can therefore be interpreted in terms 
of one or more of the remaining possibilities. 
In \S 3.3.1, we describe the observational evidences in support of a pure Kolmogorov form, and 
in \S 3.3.2, we summarize the observational data that go against it.
In \S 3.3.3, we attempt to reconcile the various observational results and discuss the possible
implications for the overall nature of the density spectrum.

%%%%%%%%%%%%%%%%%%%%%%%%%%%%%%%%%%%%%%%%%%%%%%%%%%%%%%%%%%%%%%%%%%%%%%%%%%%%%%%%%%%%%%

\subsubsection{Evidence in Favor of Kolmogorov ($ \alpha = \kolind $) Spectrum}

Our observations show that the measurements of diffractive and refractive angles, and consequently the slope 
parameter ($\beta$) derived from them, are consistent with a Kolmogorov form of spectrum (hypothesis IA of 
\S 1) for a large number (14 out of 18) pulsars. 
A quite similar result is reported by Smith \& Wright (1985), where the
measured scattering angles are found to be compatible with a Kolmogorov spectrum extending over a range of 
spatial scales from $\sim$ $ 10^9 $ m to $ \sim $ $ 10^{12} $ m. 
There are 14 pulsars common between our sample and that of Smith \& Wright (1985). 
While we find 5 pulsars with $\beta$ significantly larger than \kolind (see Table 8), the sample of 
Smith \& wright (1985) has 3 pulsars favoring $ \kolind < \alpha < 4 $ (see their Table 1).
PSR B1929+10 is an interesting case for which a steeper spectrum is suggested by both the observations 
($ \beta \approx 3.91 \pm 0.03 $ from our data and $ \refr/\diff \approx 0.61 $ from Smith \& Wright 1985).
Among the remaining 4, PSR B0834+06 is a special case with $\beta$ larger than \kolind in session II
(note that $\avbeta \approx 3.69 \pm 0.03$). For PSRs B1133+16 and B1604$-$00, no meaningful drift
measurements were made by Smith \& Wright (1985). Thus, in general, measurements of drift slopes in dynamic
spectra support a $\alpha \ \approx \ \kolind$ spectrum over the spatial scale range 
$\sim $ $ 10^7 - 10^{12} $ m.

The scaling of scintillation parameters with the frequency and/or distance is known to be sensitive to
$\alpha$.
Although there is ambiguity involved in relating the scaling exponent to $\alpha$, it can be resolved using
other observational indicators.
Cordes et al. (1985) find the frequency scaling of decorrelation bandwidths (for 5 pulsars)
to be consistent with the index $ \alpha = 3.63 \pm 0.2 $.
For the common pulsar PSR B0329+54, we find $\beta \approx 3.3 \pm 0.2$, consistent with the result of Cordes
et al. (1985). Another evidence in this direction comes from the scaling of decorrelation bandwidth and
scintillation time scale for PSR B1937+21 (Cordes et al. 1990), where the scaling implies $\alpha = 3.55 \pm
0.11$.
Further evidence in support of a Kolmogorov form comes from the VLBI observations. 
Gwinn, Moran \& Reid (1988) studied the image wander of clusters of $ {\rm H_2 O } $ masers 
in W49 and Sgr B2 and showed $\alpha \approx 3.67$ up to length scales $ \sim $ $10^{11}$ m.
Observations of the scattering disk of the pulsar PSR B1933+16 are found to be consistent 
with $\alpha = 3.52 \pm 0.13$ at length scales of $10^6$ to $10^7$ m (Gwinn et al. 1988a).

%%%%%%%%%%%%%%%%%%%%%%%%%%%%%%%%%%%%%%%%%%%%%%%%%%%%%%%%%%%%%%%%%%%%%%%%%%%%%%%%%%%%%%%%%%%%%%%%

\subsubsection{Evidence in Favor of non-Kolmogorov Forms of Spectra}

Evidence against a simple Kolmogorov form comes from a number of observations.
Many pulsars are known to show flux modulations well in excess of the Kolmogorov predictions
(see \S 3.1.1 for a discussion and summary).
Though an immediate interpretation is that the spectrum needs to be steeper (hypothesis II of \S 1), 
Kolmogorov spectrum truncated at a sufficiently large inner scale (hypothesis IB of \S 1) can also 
explain large flux modulations (cf. Coles et al. 1987; Goodman et al. 1987). 
Similarly, measurements of modulation indices of \nd and \td are found to be significantly larger 
than the Kolmogorov predictions for almost all pulsars (see \S 3.1.1), but are consistent with the 
predictions for $ \kolind < \alpha < 4.3 $ given by Romani et al. (1986). 

Further evidence favoring a {\it non-Kolmogorov} form of spectrum comes from measurements of angular 
broadening and studies of long-term DM variability.
Using the first method, Spangler \& Cordes (1988) measure $\alpha = 3.79 \pm 0.05$ towards the compact source 
2013+370, and Wilkinson, Spencer \& Nelson (1988) obtain $\alpha = 3.85 \pm 0.05$ towards Cygnus X-3.
Phillips \& Wolszczan (1991) studied DM variations of PSRs B0823+26, B0834+06 and B0919+06 over a time span
$\sim$ 2 yr, and their structure function analysis shows $\alpha$ to be larger than \kolind ($\avalpha =
3.84 \pm 0.02$) over the spatial scale range $ \sim $ $ 10^7 - 10^{13} $ m. 
At first sight, this might appear contrary to our observations, since we find the $ \beta $ values for
these pulsars to be consistent with \kolind within the errors (Table 8). 
It is important to note that our $ \beta $ values are sensitive to fluctuations on spatial scales 
$\sim$ $ 10^7 - 10^{11} $ m, whereas the DM fluctuations probe $ \sim $ $ 10^{11} - 10^{13} $ m. 
As mentioned in \S 2.3.1, from persistent drifting features observed in our PSR B0834+06 data,
we infer a similar enhancement in power level at spatial scales $ \sim $ $ 10^{12} - 10^{14} $ m
compared to the Kolmogorov expectations (see Fig. 5.a). 

Backer et al. (1993) studied DM variations of 4 pulsars (PSRs B1821$-$24, B1855+09, B1937+21 and B1951+32)
spanning the DM range 13$-$119 \dmu. They find the \delDM$-$DM relation to be much flatter than that
expected based on DISS, which they interpret as an evidence against a direct link between the density
fluctuations responsible for DM variations and those which cause DISS. They suggest a model comprising
several wedge-like structures randomly distributed along the line of sight to account for the observed DM
variations. 
It may be mentioned that Rickett et al. (1997) suggest a similar explanation for the fringing event seen 
in the PSR B0834+06 data at 408 MHz.
In both cases, the models proposed for the density fluctuations are in accordance with the hypothesis (IV)
of \S 1.

Observations of unusual scattering phenomena such as ESEs, multiple imaging events and persistent drifting
features can be considered to be a strong evidence in favour of non-Kolmogorov forms of spectra. 
Multiple imaging events are expected to be rare for type IA spectrum, but can be more common if the
spectrum is type II with $ \alpha \ \ga \ 4 $ (Hewish et al. 1985; Rickett 1990).
But the existing observational data are not adequate to make a firm statement on the statistics of their
occurrence.
The more commonly favored explanation is in terms of refraction through discrete structures 
(e.g. Cordes \& Wolszczan 1986). 
Observations of ESEs are also explained in terms of large-scale refractors ($ \sim $ 10 AU) 
in the ISM (Romani et al. 1987; Fiedler et al. 1994; Clegg, Fey \& Lazio 1998).
As discussed in \S 2.3.2, persistent drifting bands can also be understood in terms of discrete density
structures in the ISM.
Relatively rare occurrences of the three effects indicate that these density structures are localized.
As stated in \S 3.1.2, there is some evidence for the connection between these effects; for example, 
three of our pulsars $-$ PSRs B0834+06, B1133+16 and B1919+21 $-$ are known to have shown both multiple 
imaging and persistent drifts.
Thus the data accumulated so far clearly signify the importance of discrete structures in the ISM, 
thereby supporting hypothesis (IV), at least along some lines of sight. 

%%%%%%%%%%%%%%%%%%%%%%%%%%%%%%%%%%%%%%%%%%%%%%%%%%%%%%%%%%%%%%%%%%%%%%%%%%%%%%%%%%%%%%%%%%%%%%%%

\subsubsection{Overall Nature of the Density Spectrum in the Local (1 kpc) ISM}

Various methods discussed in \S 3.3.1 and \S 3.3.2 probe different parts of the density fluctuation spectrum, 
and it is interesting to see that the measurements based on a particular method give similar implications for
the nature of the spectrum.
It is worth examining what extent these observational results can be reconciled and 
what they mean for the overall nature of the spectrum.
The frequency scaling of \nd basically probes spatial scales in the range $ \sim $ $ 10^6 - 10^8 $ m, where
$ \alpha $ is found to be consistent with $ \kolind $.
VLBI angular broadening of PSR B1933+16 also probes a similar range ($ \sim $ $ 10^6 - 10^7 $ m),
and the value of $ \alpha $ inferred from this is also consistent with $ \kolind $.
The measurements of drift slopes probe spatial scales near $ \sim $ $ 10^{10} - 10^{11} $ m, and 
support $ \alpha \approx \kolind $ spectrum towards a number of lines of sight. 
Further, observations of image wander of $ {\rm H_2 O } $ masers, which also probe spatial scales of 
similar range ($ \sim $ $ 10^{11} $ m), gives another independent evidence for $ \alpha = \kolind $ spectrum.
Thus the density spectrum seem to be a power-law with the Kolmogorov index (hypothesis IA) in the range
$ \sim $ $ 10^6 - 10^{11} $ m.

Long-term DM variability, as reported by Phillips \& Wolszczan (1991), probes much larger spatial scales
($ \sim $ $ 10^{11} - 10^{13} $ m) and the results indicate that the strength of density fluctuations at
these scales is significantly larger than the Kolmogorov expectations.
As discussed in \S 2.3.1, persistent drift slopes observed in our data are also suggestive of excess
power at $ \sim $ $ 10^{12} - 10^{13} $ m.
These, combined with the results for smaller spatial scales ($ \sim $ $ 10^6 $ m to $ \sim $ $ 10^{11} $ m) 
would warrant a ``multi-component'' spectrum ($ie,$ hypothesis III of \S 1) in the range 
$ \sim $ $ 10^6 - 10^{13} $ m, with either a break near $ \kappa \ \sim \ 10^{-11} \ {\rm m ^{-1} } $ 
(type IIIA) or a ``bump'' at $ \kappa \ \sim \ 10^{-12} - 10^{-13} \ {\rm m ^{-1} } $ (type IIIB).

The modulations of DISS observables (\nd and \td) and flux, which are indicative of the strength of density 
fluctuations near refractive scales ($ \sim $ $ 10^{10} - 10^{11} $ m), are not in agreement with the 
Kolmogorov predictions (\S 2.1, \S 3.1).
However, the implications for the density spectrum here are not unambiguous
since the observed discrepancies can, at least partly, be attributed to the 
inadequacies of the thin-screen models.
The thin screen is not likely to be a valid approximation towards most pulsars within 1 kpc, and the
theoretical treatments need to be refined to analyze the perturbations of DISS parameters due to 
extended and/or inhomogeneous media.
Hence the modulation indices of observables, though discrepant with the predictions of type IA spectrum, 
do not put stringent constraints on the form of the spectrum.
The results from angular broadening measurements (towards Cyg X-3 and 2013+270) and DM variations 
of Backer et al. (1993) are also not in direct contradiction with the implications from other measurements.
The lines of sight of Cyg X-3 and 2013+270 can be treated as atypical as they are characterized by
exceedingly large strengths of scattering, predominantly from the region beyond $ \sim $ 1 kpc.
The conclusion of Backer et al. (1993) $-$ wedge-like discrete structures responsible for DM
variations $-$ is largely based on observations of distant (D \ga 1 kpc) pulsars.

Observations such as ESEs, multiple imaging and persistent drifts can be interpreted in terms of discrete
density structures in the ISM (hypothesis IV of \S 1).
Multiple imaging events have been reported only for 7 pulsars, and 4 pulsars are
known to have shown persistent drift slopes (\S 3.1.2).
To date, ESEs have been identified in the data of 9 radio sources and one pulsar.
The existing observational data therefore suggest such effects are relatively rare phenomena, 
which means discrete structures may {\it not} be {\it very} common.
Hence the hypothesis IV ($ie,$ a density spectrum of irregularities with superposed 
deterministic structures) seems to be relevant only for a limited number of lines of sight.
The overall picture emerges from the above discussion is that underlying density
fluctuations can, in general, be described by hypothesis (III) ($ie,$ a Kolmogorov-like spectrum 
which either steepens or exhibits a ``bump'' in the low wavenumber range), and hypothesis (IV) applies 
to some specific lines of sight.

%%%%%%%%%%%%%%%%%%%%%%%%%%%%%%%%%%%%%%%%%%%%%%%%%%%%%%%%%%%%%%%%%%%%%%%%%%%%%%%%%%%%%%%%%%%%%%%%

\subsection{Implications for the Theoretical Models}

The simplest scenario usually considered by the theoretical models is of a thin screen scattering 
geometry and a density irregularity spectrum of type IA ($ie, \ \sinn \ll \sdiff, \ \sout \gg \sref $).
Though the estimates of density spectral slope ($\beta$) from our data are consistent with 
$\kolind \ \la \ \alpha < 4$, the corresponding theoretical predictions for the modulation indices of 
\nd, \td and F do not match with the observations.
This inconsistency gives a clear indication of the inadequacy of theoretical models based on 
such a simple scenario.

Not many investigations have been made so far to understand refractive scintillation effects due to more 
complex, but realistic scenarios such as an extended scattering medium. 
Nevertheless, a partial reconciliation of our results is possible if we consider a geometry 
with the scattering material uniformly distributed along the line of sight.
Such a scenario in combination with $\kolind \le \alpha < 4$ will suffice to account for the measured
flux modulation indices from our data. 
If such models can also give rise to $ \sim $ 2 times larger modulations of \nd and \td compared to the
thin screen, then it is possible to reconcile the observed modulations of all the 3 quantities and
estimates of slope with $\kolind \le \alpha < 4$ ($ie,$ type IA and II spectra).
At present, theories for extended media $-$ homogeneous and/or inhomogeneous $-$ are not fully developed, 
but detailed treatments have become necessary in the light of new results from our observations.

Alternatives such as models based on hypothesis IB (say, with $ \sref > \sinn > \sdiff $) 
deserve some consideration here, as they can give rise to some observable effects akin to 
those produced by type II spectra. 
For example, such models can give rise to large modulations of flux and periodic patterns 
in dynamic spectra (cf. Goodman et al. 1987; Coles et al. 1987). 
However, at present, there is no compelling observational evidence suggesting a large inner scale.
Furthermore, it is unclear whether such models can also give rise to large modulations of 
\nd and \td, and still be consistent with $\kolind \le \beta < 4$.
In this context, models based on hypotheses (IIIA) and (IIIB) also need to be examined critically.

In addition to general phenomena such as modulations of DISS observables and flux density, and drifting
bands in dynamic spectra, a satisfactory model also needs to account for relatively rare phenomena such 
as persistent drifts, multiple imaging events and extreme scattering events.
All three effects can be understood in terms of refractive effects due to large-scale deterministic 
structures ($ie,$ hypothesis IV of \S 1), suggesting that an interconnection between them is likely.
Despite the progress made so far, we still lack quantitative interpretations describing these
phenomena and further investigations are needed.

%%%%%%%%%%%%%%%%%%%%%%%%%%%%%%%%%%%%%%%%%%%%%%%%%%%%%%%%%%%%%%%%%%%%%%%%%%%%%%%%%%%%%%%%%%%%%%%%

\section{Conclusions}

We have attempted to constrain the power spectrum of plasma density fluctuations in the ISM by studying
refractive effects in pulsar scintillation.
We have used the data from a long-term scintillation study of 18 pulsars.
Reliable and accurate estimates of diffractive and refractive scintillation properties were obtained 
by monitoring the dynamic scintillation spectra at a number of epochs spanning several months.
We studied two important and easily observable effects due to refractive scintillation: (i) modulations of
scintillation observables and flux density, and (ii) drifting bands in dynamic spectra, which provide two
independent means of constraining the form of the density irregularity spectrum.
We have considered a set of hypotheses to describe the possible potential forms of the density spectrum and 
tested them using our data.
The relevant hypotheses are: 
(IA) $ \alpha = \kolind $ (Kolmogorov spectrum), (IB) $ \alpha = \kolind $ with large inner scale; 
(II) $ \alpha > \kolind $ (`steep' spectrum); 
(IIIA) `piece-wise' power-law, (IIIB) power-law with low wavenumber enhancement; 
and (IV) power-law with superposed discrete structures.
At present, quantitative predictions are available only for the cases covered under (IA) and (II).
On comparing the observed modulation indices of diffractive scintillation observables 
$-$ decorrelation bandwidth (\nd) and scintillation time scale (\td) $-$ and pulsar flux density (F) 
with the predictions, we find that the measured values are considerably larger than the predicted values 
for a thin-screen model with a density spectrum of type IA.
The measured modulation indices are spread over a wide range of values, and are consistent with the 
predictions for power-law spectra with $\kolind < \alpha < 4.3$ (hypothesis II).
The flux density modulations will also be consistent with a smaller range $\kolind \le \alpha < 4$, 
if an extended scattering geometry with uniformly distributed scattering material along the line-of-sight 
is considered. 
Predictions are not available for the modulations of \nd and \td for such a medium.
Estimates of density spectral slope ($\beta$) are obtained from our measurements of diffractive and 
refractive scattering angles, and are found to be reasonably close to 11/3 (within the measurement 
uncertainties) for a number of pulsars (14 out of 18).
For several nearby pulsars (distance $ \sim $ 200 to 700 pc), 
$ \beta $ is found to be significantly larger than 11/3 (but less than 4). 
Thus, there are conflicting interpretations, and the results from the two methods are {\it not fully} 
reconcilable within the framework of theoretical models based on hypotheses (IA) and (II).
Further, the observations of persistent drifting bands lasting over many months seen 
in our data (e.g. PSRs B0834+06 and B1919+21) indicate that there is excess power at 
spatial scales $ \sim $ 10$-$100 AU, much larger than the refractive scales.
This would mean the spectrum either needs to be piece-wise power-law or has a bump in 
the low wavenumber range (hypotheses IIIA and IIIB respectively).
An alternative interpretation is the existence of large-scale ($\sim$ 40$-$70 AU) high 
density (\Nele $\sim$ 2$-$4 \cmc) clouds along some lines of sight (hypothesis IV).

A careful consideration of all available results from the literature and our current work leads us to the
picture of a Kolmogorov-like spectrum ($ \alpha \ \approx \ \kolind $) in the wavenumber range 
$ {\rm \sim 10^{-6} \ m^{-1} \ to \ \sim 10^{-11} \ m^{-1} } $, that either steepens or has a bump
of enhanced power at low wavenumbers ($ \kappa \ \sim \ 10^{-12} - 10^{-13} \ {\rm m ^{-1} } $). 
In addition, observations of relatively rare phenomena such as persistent drift slopes, ESEs and multiple
imaging events suggest the existence of localized high density structures along some lines of sight.
Thus the observational data indicate that the electron density fluctuations in the ISM can, in general, 
be described by hypothesis (III), and hypothesis (IV) applies to some specific lines of sight.
Unlike the case with hypotheses (I) and (II), refractive scintillation effects due to scattering media
described by hypotheses (III) and (IV) have not been fully developed.
We hope the present work will stimulate detailed theoretical works necessary towards an improved
understanding of refractive scintillation effects in pulsar signals and the power spectrum of plasma
density fluctuations in the ISM.

%%%%%%%%%%%%%%%%%%%%%%%%%%%%%%%%%%%%%%%%%%%%%%%%%%%%%%%%%%%%%%%%%%%%%%%%%%%%%%%%%%%%%%%%%%%%%%%%%%%%%%%%

{\it Acknowledgments:}
The authors wish to thank J. Chengalur and M. Vivekanand for reading an earlier version of 
this manuscript and giving useful comments.
We thank an anonymous referee for an illuminating review, which stimulated several lively discussions 
among us and also helped us in improving upon the clarity as well as the contents of the paper.

%%%%%%%%%%%%%%%%%%%%%%%%%%%%%%%%%%%%%%%%%%%%%%%%%%%%%%%%%%%%%%%%%%%%%%%%%%%%%%%%%%%%%%%%%%%%%%%%%%%%%%%%

\clearpage

\begin{appendix}

\section{Statistical Reliability of the Data}

The statistical quality of our data largely depends on the number of independent measurements (\nep)
and the number of refractive cycles (\nref) spanned during the time span of observation.
For the latter, we need to know the time scale of fluctuations of our observables.
Expectations based on simple models are that the fluctuations of all the 3 quantities 
$-$ decorrelation bandwidth (\nd), scintillation time scale (\td) and flux density (F) $-$
occur over refractive time scales (\tref), which are expected to be days to weeks at our observing 
frequency.
A structure function analysis was attempted for determining the time scales, but did not yield 
meaningful results owing to limited number of measurements.
However, since our measurements of decorrelation bandwidth and scintillation time scale are fairly accurate
(with typical uncertainties $ \sim $ $5-10$\%), first order estimates of refractive time scales
can be estimated using the expression (Rickett 1990)

\begin{equation}
\tref \approx \left( { 2 ~ \fobs \over \nd } \right) ~ \td
\end{equation}

\noindent
where \fobs is the observing frequency.
The above relation is based on simple models, i.e., scattering due to a thin screen with a power-law form of
density spectrum.
We use the estimates of \ndg and \tdg obtained from the Global ACF analysis (see Paper I) to estimate \tref.
Our values of \tref and \nref (given by \tsp/\tref, where \tsp is the time span of observation) are listed in
columns (6) and (7) of Table 3.
We do not have any pulsar for which the expected time scale of fluctuation is larger than the time span of
observation.

On the basis of the estimates of \nref, the data are divided into 3 broad categories: A ($ \nref \ge 10 $),
B ($ \nref \sim 5-10 $) and C ($ \nref < 5 $).
In a similar way, a categorization is made based on the number of measurements: A ($ \nep \ge 20 $),
B ($ \nep \sim 10-20 $) and C ($ \nep < 10 $).
These categories are listed in columns (8) and (9) respectively of Table 3.
The data which have `C' for either of the two categories, are considered to be of poor statistical quality.
These include PSRs 1540$-$06, 2016+28 and 2310+42 which have only a few cycles of fluctuations (mainly 
due to their low space velocities), data from the initial session of PSR B1133+16 (\nep = 6) 
and PSRs 1237+25, 1508+55 and 1929+10.
From Table 3, we find 7 data sets in category A (both in terms of \nep and \nref), and 11 with reasonably good
statistical reliability.
These 18 data sets are used while comparing our results with the predictions.

%%%%%%%%%%%%%%%%%%%%%%%%%%%%%%%%%%%%%%%%%%%%%%%%%%%%%%%%%%%%%%%%%%%%%%%%%%%%%%%%%%%%%%%%%%%%%%%%
\newpage

\section{Non-ISS Contributions to the Modulation Indices}

{\bf (a) Measurement Noise:} First of all, we consider the effect due to various sources of noise involved in
the measurement process.
These include (i) errors due to the Gaussian fitting done to the ACF (\smod) $-$ relevant for \nd and \td,
(ii) ``DISS noise'' or estimation errors due to the finite number of scintles in the dynamic spectrum
($\sest$) $-$ relevant for \nd, \td and F, and (iii) errors due to the flux calibration (\scal)
$-$ relevant for F.
The techniques for estimation of these quantities are discussed in detail in Paper I.
The time series of \nd, \td and F (see Figs. 4(a)$-$(x) of Paper I) give some idea about these noise sources,
where the uncertainties in \nd and \td are given by $ \sqrt { \vsmod + \vsest } $ and that in F is given
by $ \sqrt{ \vscal + \vsest } $.
The effect of these noise sources is an apparent increase in the modulations, and therefore the
measured modulation indices (\mb, \mt and \mr) need to be corrected for this.
We estimate the noise modulation indices (\mnoise) as the typical fractional uncertainties of these
quantities and these are given in columns (3), (4) and (5) of Table 4.
The measured modulation index (\mmeas) is then given by

\begin{equation}
(m _{meas}) ^2  ~ = ~ (m _{riss}) ^2 ~ + ~ (m _{noise}) ^2
\end{equation}

\noindent
The RISS-induced modulations (\mriss) of \nd, \td and F obtained in this manner are given in columns
(6), (7) and (8) of Table 4.
Since noise modulation indices are typically 0.1 for our data, their contributions to the measured
modulation indices are usually marginal.
Only exceptions are \td modulations of \normb and PSR B1929+10, for which \mnoise is slightly larger
than the estimate of \mt, and the flux modulation of \eleva with $ \mnoise \approx \mr $.
Note that the last two data are of poor statistics due to limited number of measurements ($ \nep < 10 $)
as described earlier.
Further, for part of the data of PSR B0834+06 (II and III) and PSR B2310+42, corrected estimates of
\mt are significantly lower ($ < 0.1 $) compared to their uncorrected values.
Excluding these 6 measurements and the data with poor statistical reliability, global averages
of modulation indices are: \avmb = 0.36, \avmt = 0.17 and \avmr = 0.44, very close to those obtained
from the direct measurements.

{\bf (b) Effect of variable Faraday rotation on flux density modulations:} 
Since ORT is sensitive only to linearly polarized radiation with
the electric field in the North-South plane, we need to consider the apparent flux modulation index
due to epoch-to-epoch variations of Faraday rotation (due to the Earth's ionosphere).
Significant fraction of radiation from most pulsars is known to be linearly polarized and our sample
consists of pulsars with fractional linear polarization (at 400 MHz) ranging from 0.1 to 0.8.
In Table 5, column (3) gives the fraction of linearly polarized radiation (\mlin) at 400 MHz, 
and the position angle (PA) swing across the pulse profile is given in column (4). 
These are measurements reported in the literature 
(Gould 1994; Manchester \& Taylor 1977, Hamilton et al. 1977).
Rotation measures (RM) of our pulsars are listed in column (5) of Table 5.
Adopting the method described in Gupta et al. (1993), 
we have estimated the apparent flux modulation index (\mrpol) for each pulsar.
This method takes into account the differential Faraday rotations across the observing band (\Bobs) and
across the pulse profile (\Tpulse), and estimates the worst case values of \mrpol (rotation angle 
variations across \Bobs and \Tpulse are treated as approximately linear, and ionospheric contribution
to RM is assumed to be $ \sim $ 1 \rotu).
The values of \mrpol are listed in column (6) of Table 5.
The flux modulation indices (\mrriss) given in column (8) of Table 4, which are already corrected for the
contribution due to noise modulations, are further corrected for the contribution due to \mrpol, and
the new values of \mrriss are given in column (7) of Table 5.
For 10 pulsars, \mrpol \la 0.05, and therefore this effect can be ignored.
For 6 pulsars, $ \mrpol \sim 0.1 $ to 0.2, but much smaller compared to their observed flux modulation
indices and hence the effect is only marginal.
For PSRs 1237+25 and 1929+10, for which substantial fraction of radiation is linearly polarized
(with fractional linear polarizations of 0.56 and 0.79 respectively at 408 MHz), \mrpol is estimated 
to be very large ($ \sim 0.4-0.5 $).
We also note that for these pulsars, the measured values of \mr are substantially larger than that of
the rest, substantiating the effect of Faraday rotation.
On applying the correction, we get $ \mr \approx 0.42 $ for PSR B1929+10, and 
$ \mr \approx 0.55 $ for PSR B1237+25.

{\bf (c) Effect of the Earth's orbital motion on modulations of scintillation time scale:}
The scintillation pattern speed (\viss), which determines the scintillation time scale (\td), 
is predominantly due to pulsar's proper motion.
However, in the exceptional cases of pulsars with low proper motions, contributions due to the Earth's
orbital motion (\vobs) around the Sun and the bulk flow of the density irregularities (\virr) may also
turn out to be significant, which will modify the `intrinsic' fluctuations of scintillation time scale
caused by RISS.
In order to identify data where these effects may be significant, we quantify the effect of Earth's motion
as the expected fractional variation in \viss, which is computed as the ratio of change in the transverse
component of Earth's motion (\delvep) over the observing time span (\tsp) to the scintillation speed
(\viss) computed from average values of \nd and \td.
The values of \delvep and \deltvobs = \delvep / \viss are given in columns (3) and (4) of Table 6.
Estimates of \deltvobs range from 0.01 to 0.24, but for most pulsars it can be ignored in comparison to
\mt.
Only exceptions are PSRs B1540$-$06 and B1604$-$00, for which \mt values are comparable to their 
\deltvobs, and therefore modulations of their \td measurements are not very reliable.
It is not possible to get a similar estimate for the effect due to motion of the medium, but it is known to
be significantly lower in comparison to the Earth's motion and pulsar's proper motion.
Bondi et al. (1994), based on their one-year flux modulation studies of low frequency variables, argue
that \virr $ < $ 10 \velu.
Therefore we assume \virr $ \sim $ 10 \velu and estimate the expected modulation in \td due to it
as \deltvirr = \virr/\viss.
Our estimates of \deltvirr are given in column (5) of Table 6, from which one can see that the effect
is significant only for PSR B1604$-$00, for which the values of \mt and \deltvirr are comparable.
For PSRs B1540$-$06, B2016+28, B2310+42 and B2327$-$20,
though \viss $ < $ 100 \velu, \deltvirr are considerably
lower than the measurements of \mt and hence can cause only marginal increase in the RISS-induced \td
modulations.
Thus modulations of \td due to the Earth's orbital motion and/or the motion of the density irregularities are
significant only for two pulsars.
Nevertheless, neither of the effects are reflected as a large value of \mt for these pulsars.

{\bf (d) Effect of intrinsic flux variations on the flux modulation index:} 
It is generally believed that pulsar flux variations seen at time scales of days to weeks are 
due to RISS.
But if there are some intrinsic flux variations occurring over similar time scales, 
then the measured values of \mr will be overestimates of flux modulations due to RISS.
Observations so far, have not conclusively established the occurrence of such intrinsic flux variations.
Another possibility is variations over time scales intermediate between our typical durations of
observation
(2$-$3 hours) and interval between the successive measurements (1$-$2 days).
However, we do not find any compelling reason to consider such an effect.
Recent studies of flux monitoring suggest that pulsar fluxes are stable over time scales larger than
the refractive time scales (e.g. Kaspi \& Stinebring 1992).
Though the present observations show some evidence for flux variations over time scales longer than our
typical time spans of observations (see Paper I for a discussion on long-term stability of flux densities),
such effects can be ignored in the present context.
While we do not totally rule out any hitherto unrecognized form of intrinsic flux variations,
in the absence of any other information, we assume that the observed flux modulations are largely due to RISS.

{\bf (e) Modulations of decorrelation bandwidth:}
Unlike the case with the scintillation time scale (\td) and the flux density (F), there are no non-ISS
effects in our data that can cause modulations of decorrelation bandwidth (\nd).
Our data are in general free from various kinds of man-made radio noise (mainly due to the geographical 
location of the ORT).
The fraction of data corrupted by different kinds of RFI (radio frequency interference) seldom exceeds a few 
percent, and are excluded from the analysis.
Sample data presented in Figs. 1(a)$-$(h) and 3(a)$-$(m) of Paper I give some idea about typical
quality of our data.

The modulations of \nd can result from phase gradient and/or curvature effects, whereas only the latter is
relevant for \td and F.
But the theoretical treatments incorporate both the effects in predicting the modulation index of \nd.
It is possible to correct the measured \nd at a given epoch for the refraction due to the gradient effects,
as seen from the drifting features in dynamic spectra.
This issue is discussed in \S 2.2.1.
Our analysis shows that modulation indices of corrected \nd, though somewhat lower than those of
measured \nd, they are considerably larger than the Kolmogorov predictions given in Table 1.
Further, part of our data are associated with ``persistent drifts'' or non-zero values of mean refractive
angles (this aspect is discussed in \S 2.3), which go against the expectations based on simple models.
PSRs B0834+06 (excluding data from the session IV), B1919+21, B1133+16, B1604$-$00 and B2045$-$16 belong to
this category.
It is unclear, however, whether in such cases the measured modulation indices of \nd signify their true
values.
But we do not consider this to be a strong enough reason to exclude them from the present discussion.

\end{appendix}

%%%%%%%%%%%%%%%%%%%%%%%%%%%%%%%%%%%%%%%%%%%%%%%%%%%%%%%%%%%%%%%%%%%%%%%%%%%%%%%%%%%%%%%%%%%%%%%%%%%%%%%%%%%
%%%%%%%%%%%%%%%%%%%%%%%%%%%%%%%%%%%%% REFERENCES %%%%%%%%%%%%%%%%%%%%%%%%%%%%%%%%%%%%%%%%%%%%%%%%%%%%%%%%%%
%%%%%%%%%%%%%%%%%%%%%%%%%%%%%%%%%%%%%%%%%%%%%%%%%%%%%%%%%%%%%%%%%%%%%%%%%%%%%%%%%%%%%%%%%%%%%%%%%%%%%%%%%%%

\clearpage

%%%%%%%%%%%%%%%%%%%%%%%%%%%%%%%%%%%%%%%%%%%%%%%%%%%%%%%%%%%%%%%%%%%%%%%%%%%%%%%%%%%%%%%%%%%
%%%%%%%%%%%%%%%%%%%%%%%%%%%%%%% FIGURE CAPTIONS %%%%%%%%%%%%%%%%%%%%%%%%%%%%%%%%%%%%%%%%%%%
%%%%%%%%%%%%%%%%%%%%%%%%%%%%%%%%%%%%%%%%%%%%%%%%%%%%%%%%%%%%%%%%%%%%%%%%%%%%%%%%%%%%%%%%%%%

\begin{center} {\large\bf Figure Captions} \end{center}

\begin{description}

\item [Figure 1(a)$-$(x)]
Time series of drift-corrected decorrelation bandwidth \ndc. 
The uncertainties in the measurements indicate $\pm$1-$\sigma$ error estimates, which include errors due 
to gaussian model fitting and estimation errors due to finite number of scintles at a given epoch of 
observation.
The name of the pulsar and session ID (wherever needed) are given within each panel. 
The solid markers (at either ends of the panel) indicate the average estimates of the time series.

\item [Figure 2(a)$-$(x)]
Similar to Fig. 1, except that the quantity plotted is refractive scattering angle \refr. 
The dashed line corresponds to \refr = 0. 

\item [Figure 3(a)$-$(b)]
Values of $\beta$ are plotted against DM (panel a) and distance (panel b).
For PSRs B0823+26, B0834+06, B1133+16 and B1919+21, \avbeta (the average of $\beta$ from different observing 
sessions) is used.
The dotted line indicates the Kolmogorov index \kolind.
The uncertainties in $\beta$ values are at $\pm$1-$\sigma$ levels.

\item [Figure 4(a)$-$(d)]
Sample plots of the histograms of drift rates (\dtn) shown to illustrate the drift classification scheme.
The pulsar name and session ID (wherever needed) are given at the top of each panel, and the drift class is 
given at the top right corner of the panel.
The dashed vertical line corresponds to the zero drift rate.

\item [Figure 5(a)$-$(b)]
Composite density spectrum for (a) PSR B0834+06 and (b) PSR B1919+21. 
The dotted line represents the Kolmogorov ($\alpha = \kolind$) scaling and the dot-dashed line 
$\alpha = 4$.
The dashed line corresponds to the average slope ($ {\rm \langle \beta _{meas} \rangle } $)
determined by power levels at diffractive ($\sim$ $10^{-8}~m^{-1}$) and refractive 
($\sim$ $10^{-11}~m^{-1}$) wavenumbers; this is given at the top right corner of each panel.
For PSR B0834+06, $ {\rm \langle \beta _{meas} \rangle } $ = 3.66, which makes the dashed 
line to coincide with the dotted one.
The quantity given at the bottom left corner in panels (a) and (b) is the average slope 
($ {\rm \langle \beta _{steep} \rangle } $) over the wavenumber range determined by persistent 
drifts and refractive length scales.

\end{description}

%%%%%%%%%%%%%%%%%%%%%%%%%%%%%%%%%%%%%%%%%%%%%%%%%%%%%%%%%%%%%%%%%%%%%%%%%%%%%%%%%%%%%%%%%%%%%%%%%%%%%%%%%%%

\begin{thebibliography}{}

\bibitem{} Anantharamaiah, K. R., Gothoskar, P. \& Cornwell, T. J. 1994, JAA, 15, 387
\bibitem{} Armstrong, J. W., Coles, W. A., Kojima, M. \& Rickett, B. J. 1990, ApJ, 358, 692
\bibitem{} Armstrong, J. W., Rickett, B. J. \& Spangler, S. R. 1995, ApJ, 443, 209
\bibitem{} Backer, D. C., Hama, S., Van Hook, S. \& Foster, R. S. 1993, ApJ, 404, 636
\bibitem{} Bhat, N. D. R., Gupta, Y. \& Rao, A. P. 1997,
in Proceedings of the IAU Colloquium No. 166 ``The Local Bubble and Beyond'',
eds. D. Breitschwerdt, M. J. Freyberg, J. Tr\"umper,
Lecture Notes in Physics 506, 211
(Springer-Verlag)
\bibitem{} Bhat, N. D. R., Gupta, Y. \& Rao, A. P. 1998, ApJ, 500, 262
\bibitem{} Bhat, N. D. R., Rao, A. P. \& Gupta, Y. 1998, In Press (Paper I)
\bibitem{} Blandford, R. D. \& Narayan, R. 1985, MNRAS, 213, 591
\bibitem{} Bondi, M., Padrielli, L., Gregorini, L., Mantovani, F., Shapirovskaya, N. \& Spangler, S. R. 1994,
A\&A, 287, 390
\bibitem{} Clegg, A. W., Fey, A. L. \& Lazio, T. J. 1998, ApJ, 496, 253
\bibitem{} Cognard, I., Bourgois, G., Lestrade, J., Biraud, F., Aubry, D., Darchy, B. \& Drouhin, J. 1993, 
Nature, 366, 320
\bibitem{} Coles, W. A., Frehlich, R. G., Rickett, B. J. \& Codona, J. L. 1987, ApJ, 315, 666
\bibitem{} Cordes, J. M., Pidwerbetsky, A. \& Lovelace, R. V. E. 1986, ApJ, 310, 737
\bibitem{} Cordes, J. M., Weisberg, J. M. \& Boriakoff, V. 1985, ApJ, 288, 221
\bibitem{} Cordes, J. M. \& Wolszczan, A. 1986, ApJ, 307, L27
\bibitem{} Cordes, J. M. Wolszczan, A., Dewey, R. J., Blaskiewicz, M. \& Stinebring, D. R. 1990, ApJ, 349, 245
\bibitem{} Fiedler, R. L., Dennison, B., Johnston, K. J. \& Hewish, A. 1987, Nature, 326, 675
\bibitem{} Fiedler, R. L., Dennison, B., Johnston, K. J., Waltman, E. B. \& Simon, R. S. 1994, ApJ, 430, 581
\bibitem{} Goodman, J. J., \& Narayan, R. 1985, MNRAS, 214, 519
\bibitem{} Goodman, J. J., Romani, R. W., Blandford, R. D. \& Narayan, R. 1987, MNRAS, 229, 73
\bibitem{} Gould, D. M. 1994, Ph.D. thesis, University of Manchester
\bibitem{} Gupta, Y., Rickett, B. J. \& Coles, W. A. 1993, ApJ, 403, 183
\bibitem{} Gupta, Y., Rickett, B. J. \& Lyne, A. G. 1994, MNRAS, 269, 1035
\bibitem{} Gwinn, C. R., Cordes, J. M., Bartel, A., Wolszczan, A. \& Mutel, R. 1988,
in AIP Conf. Proc. No. 174 - Radiowave Scattering in the Interstellar Medium,
ed. Cordes, J. M., Rickett, B. J. \& Backer, D. C.
(New York: AIP), 106
\bibitem{} Gwinn, C. R., Moran, J. M., \& Reid, M. J. 1988,
in AIP Conf. Proc. No. 174 - Radiowave Scattering in the Interstellar Medium,
ed. Cordes, J. M., Rickett, B. J. \& Backer, D. C.
(New York: AIP), 129
\bibitem{} Gwinn, C. R., Taylor, J. H., Weisberg, J. M. \& Rawlings, L. A. 1986, AJ, 91, 338
\bibitem{} Hamilton, P. A., McCulloch, P. M., Ables, J. G. \& Komesaroff, M. M. 1977, MNRAS, 180, 1
\bibitem{} Hewish, A. 1980, MNRAS, 192, 799
\bibitem{} Hewish, A. 1992, Phil. Trans. Royal Soc., 341, 167
\bibitem{} Hewish, A., Wolszczan, A. \& Graham, D. A. 1985, MNRAS, 213, 167
\bibitem{} Higdon, J. C. 1984, ApJ, 285, 109
\bibitem{} Higdon, J. C. 1986, ApJ, 309, 342
\bibitem{} Kaspi, V. M. \& Stinebring, D. R. 1992, ApJ, 392, 530
\bibitem{} Kuz$^{,}$min, O. A. 1992, 
in Proceedings of the IAU Colloquium No. 128 
`` The Magnetospheric Structure and Emission Mechanisms of Radio Pulsars'',
eds. T. H. Hankins, J. M. Rankin \& J. A. Gil,
(Zielona Gora: Pedagogical University Press), 287
\bibitem{} LaBrecque, D. R., Rankin, J. M. \& Cordes, J. M. 1994, AJ, 108, 1854
\bibitem{} Manchester, R. N. \& Taylor, J. H. 1977, Pulsars
(San Francisco: W. H. Freeman and Company)
\bibitem{} Narayan, R. 1992, Phil. Trans. Royal Soc., 341, 151
\bibitem{} Phillips, J. A. \& Wolszczan, A. 1991, ApJ, 382, L27
\bibitem{} Rickett, B. J. 1990, ARA\&A, 28, 561
\bibitem{} Rickett, B. J., Coles, W. A. \& Bourgois G. 1984, A\&A, 134, 390
\bibitem{} Rickett, B. J., Lyne, A. G. \& Gupta, Y. 1997, MNRAS, 287, 739
\bibitem{} Roberts, J. A. \& Ables, J. G. 1982, MNRAS, 201, 1119
\bibitem{} Romani, R. W., Blandford, R. D. \& Cordes, J. M. 1987, Nature, 328, 324
\bibitem{} Romani, R. W., Narayan, R. \& Blandford, R. D. 1986, MNRAS, 220, 19
\bibitem{} Shishov, V. I. 1973, Astr. Zh., 50, 941
\bibitem{} Sieber, W. 1982, A\&A, 113, 311
\bibitem{} Simonetti, J. H., Cordes, J. M. \& Spangler, S. R. 1984, ApJ, 284, 126
\bibitem{} Smith, F. G. \& Wright, N. C. 1985, MNRAS, 214, 97
\bibitem{} Spangler, S. R. \& Cordes, J. M. 1988, ApJ, 332, 346
\bibitem{} Spangler, S. R., Eastman, W. A., Gregorini, L., et al. 1993, A\&A, 267, 213
\bibitem{} Stinebring, D. R. \& Condon, J. J. 1990, ApJ, 352, 207
\bibitem{} Taylor, J. H. \& Cordes, J. M. 1993, ApJ, 411, 674
\bibitem{} Wilkinson, P. N., Spencer, R. E. \& Nelson, R. F. 1988, 
in IAU Symp. 129,
The Impact of VLBI on Astrophysics and Geophysics, 
ed. Reid, M J. \& Moran, J. M. 
(Dordrecht:Kluwer),305
\bibitem{} Wolszczan, A. \& Cordes, J. M. 1987, ApJ, 320, L35

\end{thebibliography}
\end{document}

\begin{deluxetable}{rclccc}
\tablenum{1}
\tablewidth{0pt}
\tablecaption{Predicted Refractive Modulations$^{\dagger}$ for $ \alpha $ = $ { 11 \over 3 } $ spectrum}
\tablehead
{
        &        &                 & \multicolumn{3}{c}{Predicted Modulation Indices}                  \\
        &        &                 &                  &            &                                   \\
	&        & D   
	                           & Decor. bandwidth (\nd) &Scint. time scale (\td) & Flux Density (F)\\
        &        &                 &                        &                        &                 \\
Sl. No. & Pulsar & (pc)            &      $m_b$             & $m_t$                  & $m_r$           \\
        &        &                 &                        &                        &		       \\
 (1)    & (2)    & (3)             & (4)                    & (5)                    & 	(6) 	       \\
}
\startdata
1  & PSR B$0329+54$ & 1430   & 0.09 & 0.04 & 0.11 \\
2  & PSR B$0628-28$ & 2140   & 0.10 & 0.05 & 0.12 \\
3  & PSR B$0823+26$ &$~380^a$& 0.11 & 0.05 & 0.13 \\
4  & PSR B$0834+06$ & ~720   & 0.11 & 0.05 & 0.14 \\
5  & PSR B$0919+06$ &$2970^b$& 0.10 & 0.05 & 0.12 \\
6  & PSR B$1133+16$ & ~270   & 0.13 & 0.06 & 0.15 \\
7  & PSR B$1237+25$ & ~560   & 0.13 & 0.06 & 0.16 \\
8  & PSR B$1508+55$ & 1930   & 0.10 & 0.05 & 0.12 \\
9  & PSR B$1540-06$ & 1160   & 0.09 & 0.04 & 0.11 \\
10 & PSR B$1604-00$ & ~590   & 0.11 & 0.05 & 0.13 \\
11 & PSR B$1747-46$ & 1080   & 0.09 & 0.05 & 0.11 \\
12 & PSR B$1919+21$ & ~660   & 0.12 & 0.06 & 0.14 \\
13 & PSR B$1929+10$ & ~170   & 0.14 & 0.06 & 0.16 \\
14 & PSR B$2016+28$ & 1100   & 0.10 & 0.05 & 0.12 \\
15 & PSR B$2020+28$ & 1300   & 0.10 & 0.05 & 0.12 \\
16 & PSR B$2045-16$ & ~640   & 0.12 & 0.06 & 0.14 \\
17 & PSR B$2310+42$ & ~960   & 0.08 & 0.04 & 0.10 \\
18 & PSR B$2327-20$ & ~490   & 0.10 & 0.05 & 0.12 \\
\tablenotetext{\dagger}{Based on the theoretical expressions given by Romani, Narayan \& Blandford (1986).}
\tablenotetext{a}{Distance estimate from parallax measurements (Gwinn et al. 1986).}
\tablenotetext{b}{Lower limit on the distance estimate.}
\enddata
\end{deluxetable}

\begin{deluxetable}{cccc}
\tablenum{2}
\tablewidth{0pt}
\tablecaption{Predicted refractive modulations for steeper spectra$^1$}
\tablehead
{
    &      &      &      \\
	       & \multicolumn{3}{c}{Predicted Modulation Indices} \\
    &      &      &      \\
Spectral index & Decorrelation bandwidth & Scintillation time scale & Flux density \\
($ \alpha $) & ($ m_b $) & ($ m_t $) & ($ m_r $) \\
    &      &      &      \\
    (1) & (2) & (3) & (4)
}
\startdata
    &      &      &      \\
4   & 0.35 & 0.17 & 0.38 \\
    &      &      &      \\
4.3 & 0.57 & 0.25 & 0.55 \\
    &      &      &      \\
\tablenotetext{1}{Note: From Romani, Narayan \& Blandford (1986). 
These are only approximate predictions (correction factor $ \gamma $ is taken to be unity).}
\enddata
\end{deluxetable}

\newcommand{\nep}{\mbox{ $ {\rm N_{ep} } $ }}
\newcommand{\tsp}{\mbox { $ {\rm T_{sp} } $ }}
\newcommand{\tref}{\mbox{ ${\rm \tau _{ref}}$ }}
\newcommand{\nref}{\mbox{ $ {\rm N_{ref} } $ }}

\begin{deluxetable}{cllcccccc}
\tablenum{3}
\tablewidth{0pt}
\tablecaption{Statistical Quality and Reliability of the Data}
\tablehead
{
No. & Pulsar (Session) & Period of & \nep & \tsp & \tref & \nref & \multicolumn{2}{c}{Reliability} \\
    &        &Observation&      &(days)&(days) &       & Cycles       &	Measurements \\
    &        &           &      &      &       &       &              &              \\
    (1) & (2) & (3) & (4) & (5) & (6) & (7) & (8) & (9) \\
}
\startdata
01 & PSR B$0329+54$     & Apr 95$-$Jul 95 & 14  &  67 &   14.1 &  5 &    B  & B  \\
02 & PSR B$0628-28$     & Oct 93$-$Jan 94 & 17  &  84 &   17.0 &  5 &    B  & B  \\
03 & PSR B$0823+26$(I)  & Mar 93$-$May 93 & 12  &  67 &    3.7 & 18 &    A  & B  \\
04 & PSR B$0823+26$(II) & Oct 93$-$Jan 94 & 19  &  88 &    3.6 & 24 &    A  & B  \\
05 & PSR B$0834+06$(I)  & Jan 93$-$May 93 & 27  & 121 &    6.0 & 20 &    A  & A  \\
06 & PSR B$0834+06$(II) & Oct 93$-$Jan 94 & 18  &  75 &    6.6 & 11 &    A  & B  \\
07 & PSR B$0834+06$(III)& Feb 94$-$Jun 94 & 21  &  79 &    5.2 & 15 &    A  & A  \\
08 & PSR B$0834+06$(IV) & Apr 95$-$Jul 95 & 27  & 101 &    3.9 & 26 &    A  & A  \\
09 & PSR B$0919+06$     & Mar 94$-$Jun 94 & 19  &  71 &    4.7 & 15 &    A  & B  \\
10 & PSR B$1133+16$(I)  & Jan 93$-$Feb 93 &  6  &  20 &    1.3 & 16 &    A  & C  \\
11 & PSR B$1133+16$(II) & Feb 94$-$Jun 94 & 25  &  91 &    1.5 & 60 &    A  & A  \\
12 & PSR B$1133+16$(III)& Apr 95$-$Jul 95 & 27  &  96 &    1.1 & 86 &    A  & A  \\
13 & PSR B$1237+25$     & Oct 93$-$Jan 94 &  9  &  78 &    1.8 & 43 &    A  & C  \\
14 & PSR B$1508+55$     & Apr 95$-$Jul 95 &  9  &  53 &    6.2 &  9 &    B  & C  \\
15 & PSR B$1540-06$     & Apr 95$-$Jul 95 & 12  &  92 &   35.8 &  3 &    C  & B  \\
16 & PSR B$1604-00$     & Apr 95$-$Jul 95 & 10  &  94 &   18.6 &  5 &    B  & B  \\
17 & PSR B$1747-46$     & Apr 95$-$Jul 95 & 12  &  93 &   10.5 &  9 &    B  & B  \\
18 & PSR B$1919+21$(I)  & Mar 93$-$May 93 & 15  &  60 &    9.8 &  6 &    B  & B  \\
19 & PSR B$1919+21$(II) & Oct 93$-$Jan 94 & 48  &  88 &    4.4 & 20 &    A  & A  \\
20 & PSR B$1929+10$     & Mar 94$-$Jun 94 &  9  &  43 &    2.0 & 21 &    A  & C  \\
21 & PSR B$2016+28$     & Oct 93$-$Jan 94 & 20  &  84 &   36.5 &  2 &    C  & A  \\
22 & PSR B$2020+28$     & Mar 94$-$Jun 94 & 15  &  75 &    7.8 & 10 &    A  & B  \\
23 & PSR B$2045-16$     & Oct 93$-$Jan 94 & 35  &  86 &    1.9 & 44 &    A  & A  \\
24 & PSR B$2310+42$     & Apr 95$-$Jul 95 & 10  &  78 &   20.5 &  4 &    C  & B  \\
25 & PSR B$2327-20$     & Mar 93$-$Jun 94 & 18  &  69 &   12.2 &  6 &    B  & B  \\
\tablenotetext{}{Notes: Multiple entries of some pulsars indicate distinct observing sessions.}
\enddata
\end{deluxetable}

\newcommand{\mb}{$ m_b $}
\newcommand{\mt}{$ m_t $}
\newcommand{\mr}{$ m_r $}

\newcommand{\kolind}{\mbox{$ { 11 \over 3 } $ }}

\newcommand{\mbn}{$ m_{b,noise} $}
\newcommand{\mtn}{$ m_{t,noise} $}
\newcommand{\mrn}{$ m_{r,noise} $}
\newcommand{\mbr}{$ m_{b,riss} $}
\newcommand{\mtr}{$ m_{t,riss} $}
\newcommand{\mrr}{$ m_{r,riss} $}

\begin{deluxetable}{clcccccl}
\tablenum{4}
\tablewidth{0pt}
\tablecaption{Effect of Noise on Refractive Modulations}
\tablehead
{
        &        & \multicolumn{3}{c}{Noise Modulation Indices} 
		 & \multicolumn{3}{c}{Corrected Modulation Indices} \\
        &                  &      &      &      &      &      &      \\
 Sl No. & Pulsar (Session) & \mbn & \mtn & \mrn & \mbr & \mtr & \mrr \\
        &                  &      &      &      &      &      &      \\
(1)     & (2)              & (3)  & (4)  & (5)  & (6)  & (7)  & (8)  \\
}
\startdata
1  & PSR B$0329+54$  	& 0.11&0.12&0.11& 0.39&0.12&0.20$^*$         \\
2  & PSR B$0628-28$  	& 0.15&0.18&0.12& 0.41&0.14&0.28$^*$          \\
3  & PSR B$0823+26$(I)	& 0.13&0.15&0.11& 0.36&0.09&0.41             \\
4  & PSR B$0823+26$(II)	& 0.15&0.15&0.11& 0.32&0.00&0.35$^*$          \\
5  & PSR B$0834+06$(I)	& 0.12&0.13&0.14& 0.34&0.23&0.52             \\
6  & PSR B$0834+06$(II)	& 0.12&0.14&0.12& 0.34&0.06&0.34$^*$          \\
7  & PSR B$0834+06$(III)& 0.10&0.12&0.11& 0.18&0.05&0.34$^*$          \\
8  & PSR B$0834+06$(IV)	& 0.13&0.15&0.13& 0.36&0.21&0.33$^*$          \\
9  & PSR B$0919+06$  	& 0.09&0.10&0.11& 0.24&0.16&0.48             \\
10 & PSR B$1133+16$(I)	& 0.10&0.11&0.11& 0.27&0.17&0.18             \\
11 & PSR B$1133+16$(II)	& 0.12&0.11&0.13& 0.42&0.29&0.21$^@$         \\
12 & PSR B$1133+16$(III)& 0.12&0.11&0.12& 0.39&0.14&0.32$^*$          \\
13 & PSR B$1237+25$  	& 0.15&0.15&0.15& 0.30&0.27&0.67             \\
14 & PSR B$1508+55$  	& 0.10&0.10&0.11& 0.29&0.11&0.21             \\
15 & PSR B$1540-06$  	& 0.12&0.13&0.11& 0.23&0.14&0.48             \\
16 & PSR B$1604-00$  	& 0.13&0.13&0.14& 0.42&0.10&0.45             \\
17 & PSR B$1747-46$  	& 0.11&0.12&0.11& 0.37&0.11&0.21$^*$          \\
18 & PSR B$1919+21$(I)	& 0.11&0.12&0.13& 0.13&0.17&0.47             \\
19 & PSR B$1919+21$(II)	& 0.13&0.15&0.14& 0.33&0.12&0.41             \\
20 & PSR B$1929+10$  	& 0.19&0.19&0.19& 0.34&0.00&0.64             \\
21 & PSR B$2016+28$  	& 0.16&0.17&0.16& 0.44&0.29&0.37             \\
22 & PSR B$2020+28$  	& 0.10&0.11&0.11& 0.19&0.10&0.56             \\
23 & PSR B$2045-16$  	& 0.13&0.15&0.12& 0.38&0.13&0.41             \\
24 & PSR B$2310+42$  	& 0.13&0.12&0.11& 0.16&0.07&0.34             \\
25 & PSR B$2327-20$  	& 0.13&0.14&0.13& 0.48&0.47&0.30$^*$          \\
\tablenotetext{*}
{Reasonable agreement with the predictions of an extended 
homogeneous medium with $\alpha = \kolind$.}
\tablenotetext{@}{Much less than the Kolmogorov prediction (0.35) of an extended medium.}
\enddata
\end{deluxetable}

\begin{deluxetable}{ccccccl}
\tablenum{5}
\tablewidth{0pt}
\tablecaption{Effect of Variable Faraday Rotattion on Flux Modulations}
\tablehead
{
No. & Pulsar & $ m_{lin} $ & PA swing & RM                  & $ m_{r,pol} $ & Corrected    \\
    &        &             & (deg)    & (${\rm rad~m^{-2}}$)&               & $m_{r,riss}$ \\
    &        &             &          &                     &               &              \\
(1) & (2)    & (3)         & (4)      &(5)                  & (6)           & (7)                          
}
\startdata
01 & PSR B$0329+54$ & 0.11$^a$ & 105     & $ -63.7 $ & 0.003 & 0.20 \\
02 & PSR B$0628-28$ & 0.32$^a$ & 145     & $  46.2 $ & 0.019 & 0.28 \\
03 & PSR B$0823+26$ & 0.29$^a$ &  50     & $   5.9 $ & 0.177 & 0.37 (I) \\
   &                &          &         &           &       & 0.30 (II) \\
04 & PSR B$0834+06$ & 0.03$^a$ & 140     & $  23.6 $ & 0.005 & 0.52 (I) \\
   &                &          &         &           &       & 0.34 (II) \\
   &                &          &         &           &       & 0.34 (III) \\
   &                &          &         &           &       & 0.33 (IV) \\
05 & PSR B$0919+06$ & 0.34$^a$ &  50     & $  32.0 $ & 0.138 & 0.46 \\
06 & PSR B$1133+16$ & 0.23$^a$ &  90     & $   3.9 $ & 0.103 & 0.15 (I) \\
   &                &          &         &           &       & 0.18 (II) \\
   &                &          &         &           &       & 0.30 (III) \\
07 & PSR B$1237+25$ & 0.56$^a$ &  30     & $  -0.3 $ & 0.378 & 0.55 \\
08 & PSR B$1508+55$ & 0.15$^a$ & 160$^d$ & $   0.8 $ & 0.013 & 0.21 \\
09 & PSR B$1540-06$ & 0.25$^a$ &  25     & $   4.0 $ & 0.170 & 0.45 \\
10 & PSR B$1604-00$ & 0.14$^a$ & 110     & $   6.5 $ & 0.048 & 0.45 \\
11 & PSR B$1747-46$ & 0.26$^c$ &  90     & $  18.0 $ & 0.102 & 0.18 \\
12 & PSR B$1919+21$ & 0.29$^a$ &  25     & $ -16.5 $ & 0.182 & 0.43 (I) \\
   &                &          &         &           &       & 0.37 (II) \\
13 & PSR B$1929+10$ & 0.79$^a$ &  50     & $  -6.1 $ & 0.486 & 0.42 \\
14 & PSR B$2016+28$ & 0.12$^b$ & 120     & $ -34.6 $ & 0.023 & 0.37 \\
15 & PSR B$2020+28$ & 0.40$^b$ & 120     & $ -74.7 $ & 0.009 & 0.56 \\
16 & PSR B$2045-16$ & 0.27$^a$ & 150     & $ -10.0 $ & 0.035 & 0.41 \\
17 & PSR B$2310+42$ & 0.17$^a$ & 135     & $   7.0 $ & 0.035 & 0.34 \\
18 & PSR B$2327-20$ & 0.24$^a$ & 150     & $   9.5 $ & 0.031 & 0.30 \\
\tablenotetext{}{Notes: The ionospheric contribution to RM 
is assumed to be $ {\rm \sim ~ 1 ~ rad ~ m^{-2} } $.}
\tablenotetext{}{The symbols I$-$IV indidate the observing session.}
\tablenotetext{a}{Gould (1994)}
\tablenotetext{b}{Manchester \& Taylor (1977)}
\tablenotetext{c}{Hamilton et al. (1977)}
\tablenotetext{d}{Measurement at 610 MHz}
\enddata
\end{deluxetable}

%%%%%%%%%%%%%%%%%%%%%%%%%%%%%%%%%%%%%%%%%%%%%%%%%%%%%%%%%%%%%%%%%%%%%%%%%%%%%%%%%%%%%%%
\newcommand{\viss}{\mbox{ $ V_{iss} $ }}
\newcommand{\delvep}{\mbox{ $ \Delta V_{obs _{\bot }} $}}
\newcommand{\velu}{\mbox{$ {\rm km ~ secs ^ {-1} } $}}
%%%%%%%%%%%%%%%%%%%%%%%%%%%%%%%%%%%%%%%%%%%%%%%%%%%%%%%%%%%%%%%%%%%%%%%%%%%%%%%%%%%%%%%

\begin{deluxetable}{clccc}
\tablenum{6}
\tablewidth{0pt}
\tablecaption{Modulations of scintillation time scale due to $ {\rm V_{obs} } $ and $ {\rm V_{irr} } $ }
\tablehead{
No. & Pulsar (Session)           & 
$ \delvep $                      & 
$ {\rm \delta t_{vobs} } $       & $ {\rm \delta t_{virr} ^{\dagger} } $ \\
    &                            &
    (\velu)                      &
                                 &                                       \\
   &                    &       &    &      \\
 (1) & (2) & (3) & (4) & (5) 
}
\startdata
01 & PSR B$0329+54$  	& ~6.3 &0.03& 0.05 \\
02 & PSR B$0628-28$  	& ~5.3 &0.03& 0.06 \\
03 & PSR B$0823+26$(I)	& 11.4 &0.05& 0.04 \\
04 & PSR B$0823+26$(II)	& 27.5 &0.10& 0.04 \\
05 & PSR B$0834+06$(I)	& 22.6 &0.14& 0.06 \\
06 & PSR B$0834+06$(II)	& 23.5 &0.12& 0.05 \\
07 & PSR B$0834+06$(III)& 12.2 &0.05& 0.04 \\
08 & PSR B$0834+06$(IV)	& 21.4 &0.09& 0.04 \\
09 & PSR B$0919+06$  	& 13.8 &0.02& 0.02 \\
10 & PSR B$1133+16$(I)	& ~4.0 &0.01& 0.02 \\
11 & PSR B$1133+16$(II)	& 20.6 &0.06& 0.03 \\
12 & PSR B$1133+16$(III)& 17.0 &0.04& 0.03 \\
13 & PSR B$1237+25$  	& 13.1 &0.04& 0.03 \\
14 & PSR B$1508+55$  	& ~1.5 &0.00& 0.02 \\
15 & PSR B$1540-06$  	& 18.5 &0.24& 0.13 \\
16 & PSR B$1604-00$  	& 15.1 &0.24& 0.16 \\
17 & PSR B$1747-46$  	& ~7.3 &0.03& 0.05 \\
18 & PSR B$1919+21$(I)	& ~2.6 &0.02& 0.08 \\
19 & PSR B$1919+21$(II)	& ~9.4 &0.05& 0.05 \\
20 & PSR B$1929+10$  	& ~8.6 &0.06& 0.07 \\
21 & PSR B$2016+28$  	& ~7.5 &0.14& 0.18 \\
22 & PSR B$2020+28$  	& ~3.8 &0.02& 0.05 \\
23 & PSR B$2045-16$  	& 29.2 &0.05& 0.02 \\
24 & PSR B$2310+42$  	& ~3.7 &0.04& 0.11 \\
25 & PSR B$2327-20$  	& 17.4 &0.22& 0.13 \\
\tablenotetext{\dagger}{Assuming $ {\rm V_{irr} } $ $ \sim $ 10 \velu (Bondi et al. 1994).}
\enddata
\end{deluxetable}

\newcommand{\dtnu}{$ {\rm secs ~ kHz ^ {-1} } $}

\begin{deluxetable}{clccccccc}
\tablenum{7}
\tablewidth{0pt}
\tablecaption{Properties of drift rates and corrected bandwidths}
\tablehead
{
No. & Pulsar (Session) & $ \langle d t / d \nu \rangle $ & $ \delta ( d t / d \nu ) $ & Drift Class &
$ \langle \nu _{d_c} \rangle $ & $ m_{b,c} $ & $ m_{noise} $ & $ m_{b,c(riss)} $ \\	
    &        & (\dtnu) & (\dtnu) & 		&	(kHz) 	&	&	&	\\
    (1) & (2) & (3) & (4) & (5) & (6) & (7) & (8) & (9)
}
\startdata
01& PSR B$0329+54$     & $ -0.09 $ $ \pm $ 0.04 & 0.10 $ \pm $ 0.06 &  NC &~219&0.28 &0.11 &0.26 \\
02& PSR B$0628-28$     & $ +0.04 $ $ \pm $ 0.08 & 0.27 $ \pm $ 0.11 &  I  &~307&0.48 &0.13 &0.46 \\
03& PSR B$0823+26$(I)  & $ +0.00 $ $ \pm $ 0.02 & 0.15 $ \pm $ 0.02 &  I  &~452&0.28 &0.13 &0.24 \\
04& PSR B$0823+26$(II) & $ -0.06 $ $ \pm $ 0.02 & 0.08 $ \pm $ 0.03 &  NC &~338&0.27 &0.16 &0.22 \\
05& PSR B$0834+06$(I)  & $ +0.40 $ $ \pm $ 0.01 & 0.16 $ \pm $ 0.02 &  II &~716&0.33 &0.13 &0.30 \\
06& PSR B$0834+06$(II) & $ +0.39 $ $ \pm $ 0.02 & 0.34 $ \pm $ 0.03 &  II &~504&0.23 &0.12 &0.19 \\
07& PSR B$0834+06$(III)& $ +0.29 $ $ \pm $ 0.01 & 0.18 $ \pm $ 0.02 &  II &~488&0.25 &0.11 &0.23 \\
08& PSR B$0834+06$(IV) & $ +0.08 $ $ \pm $ 0.01 & 0.18 $ \pm $ 0.02 &  I  &~696&0.35 &0.12 &0.33 \\
09& PSR B$0919+06$     & $ +0.01 $ $ \pm $ 0.02 & 0.14 $ \pm $ 0.02 &  I  &~305&0.21 &0.10 &0.19 \\
10& PSR B$1133+16$(I)  & $ +0.01 $ $ \pm $ 0.01 & 0.06 $ \pm $ 0.01 &  I  &~554&0.23 &0.12 &0.20 \\
11& PSR B$1133+16$(II) & $ +0.03 $ $ \pm $ 0.01 & 0.09 $ \pm $ 0.01 &  I  &~986&0.39 &0.14 &0.36 \\
12& PSR B$1133+16$(III)& $ -0.05 $ $ \pm $ 0.01 & 0.06 $ \pm $ 0.01 &  NC &1763&0.36 &0.16 &0.32 \\
13& PSR B$1237+25$     & $ +0.04 $ $ \pm $ 0.01 & 0.06 $ \pm $ 0.01 &  NC &1241&0.50 &0.16 &0.47 \\
14& PSR B$1508+55$     & $ +0.01 $ $ \pm $ 0.03 & 0.07 $ \pm $ 0.04 &  I  &~268&0.27 &0.10 &0.26 \\
15& PSR B$1540-06$     & $ +0.22 $ $ \pm $ 0.12 & 0.27 $ \pm $ 0.17 &  I  &~195&0.10 &0.12 &....$^a$ \\
16& PSR B$1604-00$     & $ +0.63 $ $ \pm $ 0.08 & 0.91 $ \pm $ 0.11 &  I  &~508&0.53 &0.15 &0.50 \\
17& PSR B$1747-46$     & $ +0.04 $ $ \pm $ 0.04 & 0.12 $ \pm $ 0.06 &  I  &~231&0.27 &0.11 &0.25 \\
18& PSR B$1919+21$(I)  & $ +0.39 $ $ \pm $ 0.03 & 0.26 $ \pm $ 0.04 &  II &~424&0.15 &0.13 &0.07$^b$ \\
19& PSR B$1919+21$(II) & $ +0.17 $ $ \pm $ 0.01 & 0.17 $ \pm $ 0.01 &  II &~626&0.30 &0.13 &0.27 \\
20& PSR B$1929+10$     & $ +0.08 $ $ \pm $ 0.01 & 0.19 $ \pm $ 0.02 &  I  &1444&0.36 &0.20 &0.30 \\
21& PSR B$2016+28$     & $ +0.30 $ $ \pm $ 0.14 & 0.79 $ \pm $ 0.20 &  I  &~284&0.32 &0.16 &0.28 \\
22& PSR B$2020+28$     & $ -0.10 $ $ \pm $ 0.03 & 0.21 $ \pm $ 0.05 &  I  &~272&0.22 &0.09 &0.20 \\
23& PSR B$2045-16$     & $ +0.06 $ $ \pm $ 0.01 & 0.09 $ \pm $ 0.01 &  NC &~671&0.33 &0.11 &0.32 \\
24& PSR B$2310+42$     & $ -0.18 $ $ \pm $ 0.09 & 0.17 $ \pm $ 0.12 &  NC &~190&0.06 &0.11 &....$^a$ \\
25& PSR B$2327-20$     & $ -0.04 $ $ \pm $ 0.07 & 0.96 $ \pm $ 0.10 &  I  &~328&0.36 &0.14 &0.33 \\
\tablenotetext{a}{Noise modulations dominate over the observed fluctuations of $ \nu _{d_c} $.}
\tablenotetext{b}{The observed fluctuations of $ \nu _{d_c} $ are comparable to the noise modulations.}
\enddata
\end{deluxetable}

\newcommand{\refmean}{\mbox{$ {\rm \langle \theta _{ref} \rangle } $ }}
\newcommand{\refrms}{\mbox{$ {\rm \delta \theta _{ref} } $}}
\newcommand{\diffmean}{\mbox{ $ {\rm \langle \theta _{diff} \rangle } $ }}

\begin{deluxetable}{clcccl}
\tablenum{8}
\tablewidth{0pt}
\tablecaption{Estimates of Diffractive and Refractive Angles and Spectral Slopes}
\tablehead
{
No. & Pulsar (Session) & \refmean & \refrms$^{\dagger}$ & \diffmean &  \multicolumn{1}{c}{$\beta$} \\
    &        & (mas)    & (mas)   & (mas)     &  \\
        &     &     &     &     &         \\
    (1) & (2) & (3) & (4) & (5) & \multicolumn{1}{c}{(6)} 
}
\startdata
01& PSR B$0329+54$      & $-0.02\pm0.02$ & 0.04$\pm$0.03& 0.73$\pm$0.01& 3.29$\pm$0.21 \\
02& PSR B$0628-28$      & $+0.01\pm0.01$ & 0.07$\pm$0.03& 0.51$\pm$0.01& 3.50$\pm$0.10 \\
03& PSR B$0823+26$(I)   & $+0.01\pm0.03$ & 0.34$\pm$0.05& 0.95$\pm$0.02& 3.73$\pm$0.04 \\
04& PSR B$0823+26$(II)  & $-0.10\pm0.03$ & 0.18$\pm$0.06& 1.09$\pm$0.02& 3.54$\pm$0.09 \\
05& PSR B$0834+06$(I)   & $+0.24\pm0.01$ & 0.12$\pm$0.01& 0.54$\pm$0.01& 3.58$\pm$0.03 \\
06& PSR B$0834+06$(II)  & $+0.25\pm0.01$ & 0.26$\pm$0.03& 0.63$\pm$0.01& 3.77$\pm$0.03${\ddagger}$ \\
07& PSR B$0834+06$(III) & $+0.23\pm0.01$ & 0.20$\pm$0.02& 0.64$\pm$0.01& 3.69$\pm$0.03 \\
08& PSR B$0834+06$(IV)  & $+0.07\pm0.01$ & 0.19$\pm$0.02& 0.54$\pm$0.01& 3.71$\pm$0.03 \\
09& PSR B$0919+06$      & $+0.00\pm0.01$ & 0.10$\pm$0.02& 0.41$\pm$0.01& 3.65$\pm$0.04 \\
10& PSR B$1133+16$(I)   & $+0.00\pm0.04$ & 0.33$\pm$0.07& 0.96$\pm$0.02& 3.70$\pm$0.06 \\
11& PSR B$1133+16$(II)  & $+0.06\pm0.01$ & 0.36$\pm$0.03& 0.75$\pm$0.01& 3.78$\pm$0.02${\ddagger}$ \\
12& PSR B$1133+16$(III) & $-0.15\pm0.01$ & 0.29$\pm$0.02& 0.55$\pm$0.01& 3.80$\pm$0.02${\ddagger}$ \\
13& PSR B$1237+25$      & $+0.05\pm0.01$ & 0.12$\pm$0.03& 0.48$\pm$0.01& 3.51$\pm$0.08 \\
14& PSR B$1508+55$      & $-0.00\pm0.02$ & 0.06$\pm$0.03& 0.56$\pm$0.01& 3.44$\pm$0.15 \\
15& PSR B$1540-06$      & $+0.03\pm0.02$ & 0.06$\pm$0.05& 0.86$\pm$0.01& 3.39$\pm$0.18 \\
16& PSR B$1604-00$      & $+0.14\pm0.02$ & 0.33$\pm$0.04& 0.72$\pm$0.01& 3.79$\pm$0.03${\ddagger}$ \\
17& PSR B$1747-46$      & $+0.02\pm0.02$ & 0.08$\pm$0.05& 0.80$\pm$0.01& 3.45$\pm$0.13 \\
18& PSR B$1919+21$(I)   & $+0.20\pm0.01$ & 0.20$\pm$0.03& 0.73$\pm$0.01& 3.66$\pm$0.04 \\
19& PSR B$1919+21$(II)  & $+0.12\pm0.01$ & 0.16$\pm$0.01& 0.59$\pm$0.01& 3.63$\pm$0.03 \\
20& PSR B$1929+10$      & $+0.14\pm0.03$ & 0.57$\pm$0.05& 0.76$\pm$0.02& 3.91$\pm$0.03${\ddagger}$ \\
21& PSR B$2016+28$      & $+0.04\pm0.02$ & 0.14$\pm$0.03& 0.72$\pm$0.01& 3.59$\pm$0.06 \\
22& PSR B$2020+28$      & $-0.05\pm0.01$ & 0.11$\pm$0.03& 0.66$\pm$0.01& 3.54$\pm$0.07 \\
23& PSR B$2045-16$      & $+0.10\pm0.01$ & 0.19$\pm$0.02& 0.58$\pm$0.01& 3.69$\pm$0.03 \\
24& PSR B$2310+42$      & $-0.05\pm0.03$ & 0.06$\pm$0.06& 0.96$\pm$0.02& 3.38$\pm$0.20 \\
25& PSR B$2327-20$      & $-0.02\pm0.03$ & 0.42$\pm$0.05& 0.98$\pm$0.01& 3.78$\pm$0.03${\ddagger}$ \\
\tablenotetext{\dagger}{Measured values scaled by $ \sqrt{2} $, as described in the text (\S 2.2.2).} 
\tablenotetext{\ddagger}{Estimates that are significantly larger than $ { 11 \over 3 } $.} 
\enddata
\end{deluxetable}

\newcommand{\refkol}{\mbox{$ {\rm \delta \theta _{r,kol} } $ }}
\newcommand{\avvissc}{\mbox{$ \langle V_{iss,c} \rangle $ }}
\newcommand{\Nele}{\mbox{$ {\rm N_e } $\,}}
\newcommand{\cmc}{$ {\rm cm ^ {-3} } $ }

\begin{deluxetable}{lccccrc}
\tablenum{9}
\tablewidth{0pt}
\tablecaption{Constraints on the characteristics of high density clouds}
\tablehead
{
\multicolumn{1}{c}{Data} & \refmean & \avvissc & \tsp   & \Nele  & S$^{\dagger}$  & \refkol \\
                         & (mas)    & (\velu)    & (days) & (\cmc)   & (AU) & (mas) \\ \\
\multicolumn{1}{c}{(1)}  & (2) & (3) & (4) & (5) & (6) & (7)
}
\startdata
                         &                   &     &     &      &      &      \\
PSR B$0834+06$(I)        & 0.24 $ \pm $ 0.01 & 217 & 120 & 3.6  & 15   & 0.09 \\
PSR B$0834+06$(II)       & 0.25 $ \pm $ 0.02 & 228 & 75  & 3.7  & 10   & 0.11 \\
PSR B$0834+06$(III)      & 0.23 $ \pm $ 0.01 & 270 & 75  & 3.4  & 12   & 0.11 \\
                         &                   &     &     &      &      &      \\
PSR B0834+06$^{\ddagger}$& 0.24              & 238 & 500 & 3.6  & 70   & 0.07 \\
                         &                   &     &     &      &      &      \\
PSR B$1919+21$(I)        & 0.20 $ \pm $ 0.01 & 160 & 60  & 2.9  & 6    & 0.13 \\
PSR B$1919+21$(II)       & 0.12 $ \pm $ 0.01 & 226 & 90  & 1.8  & 12   & 0.10 \\
                         &                   &     &     &      &      &      \\
PSR B1919+21$^{\ddagger}$& 0.16              & 193 & 350 & 2.3  & 40   & 0.09 \\
                         &                   &     &     &      &      &      \\
\tablenotetext{\dagger}{Lower limit on the spatial length scale.}
\tablenotetext{\ddagger}{Combined data from the observing sessions given above.}
\enddata
\end{deluxetable}

\addtolength{\oddsidemargin}{-1.1cm}

\newcommand{\lp}{\mbox{\raisebox{-0.1ex}{$\scriptscriptstyle \stackrel{<}{\sim}$\,}}}
\newcommand{\gt}{\mbox{$ > $}}

\begin{deluxetable}{ccccrrrccc}
\tablenum{10}
\tablewidth{0pt}
\tablecaption{Flux modulation indices from the literature}
\tablehead
{
No. & Pulsar & ORT${\rm ^a}$ & GRL94${\rm ^b}$ & LRC93${\rm ^c}$ & GRC93${\rm ^d}$ & KS92${\rm ^e}$ & \multicolumn{3}{c}{SC90${\rm ^f}$} \\
 & & (327 MHz) & (408 MHz) & (408 MHz) & (74 MHz) & (610 MHz) & (310 MHz) & (416 MHz) & (750 MHz) \\
    &        &     &       &           &       &      &     &     &     \\
(1) & (2) & (3) & (4) & (5) & (6) & (7) & (8) & (9) & (10)
}
\startdata
01 & PSR B$0329+54$&  0.23 &  0.40 &          & \gt 0.15 & 0.39     & & & \\
02 & PSR B$0628-28$&  0.30 &  0.33 &          &          &          & & & \\
03 & PSR B$0823+26$&  0.40 &  0.22 & \lp 0.45 & \gt 0.16 &          & & & \\
04 & PSR B$0834+06$&  0.40 &  0.53 & \lp 0.36 & 0.16     &          & & & \\
05 & PSR B$1133+16$&  0.30 &       &          & 0.18     &          & & & \\
06 & PSR B$1237+25$&  0.69 &       &          & 0.30     &          & & & \\
07 & PSR B$1508+55$&  0.24 &       &          & 0.28     & \gt 0.11 & 0.24 & 0.25 & 0.33 \\
08 & PSR B$1604-00$&  0.47 &       & \lp 0.71 &          &          & 0.43 & 1.04 & 1.06 \\
09 & PSR B$1919+21$&  0.47 &       &          & \gt 0.21 &          & 0.69 & 0.80 & 0.97 \\
10 & PSR B$2016+28$&  0.40 &  0.25 &          &          &          & 0.18 & 0.21 & 0.35 \\
11 & PSR B$2020+28$&  0.57 &       & \lp 0.36 &          &          & 0.38 & 0.53 & 0.50 \\
12 & PSR B$2045-16$&  0.42 &       &          &          &          & 0.60 & 0.81 & 1.23 \\
\tablenotetext{}{Notes: ${\rm ^a}$Measurements from our data, ${\rm ^b}$Gupta, Rickett \& Lyne (1994), 
${\rm ^c}$LaBrecque, Rankin \& Cordes (1993), ${\rm ^d}$Gupta, Rickett \& Coles (1993), ${\rm ^e}$ Kaspi \& Stinebring (1992),
${\rm ^f}$Stinebring \& Condon (1990). The frequency of observation is given in brackets.}

\enddata
\end{deluxetable}